\documentclass[10pt,journal,compsoc]{IEEEtran}

\usepackage{caption}
\usepackage{hyperref}
\usepackage{ragged2e}
\usepackage{multirow}
\usepackage{diagbox}
\usepackage{color}
\usepackage{censor}
\usepackage{url}
\usepackage{booktabs}
\usepackage{pifont}
\usepackage{enumitem}
\newtheorem{definition}{Definition}
\newcommand{\toolname}{AG-RAG}
\usepackage{xspace}
\newcommand{\ie}{\textit{i.e.,}\xspace}
\newcommand{\eg}{\textit{e.g.,}\xspace}

\newcommand{\etal}{\textit{et al.}\xspace}

\newcommand{\figref}[1]{Fig.~\ref{#1}\xspace}

\newcommand{\newdata}{$Data_{new}$\xspace}
\newcommand{\olddata}{$Data_{old}$\xspace}

\newcommand{\irar}{$\mathit{IR_{ar}}$\xspace}
\newcommand{\rah}{$\mathit{RA_{adapt}^{H}}$\xspace}
\newcommand{\rann}{$\mathit{RA_{adapt}^{NN}}$\xspace}
\newcommand{\inte}{$\mathit{Integration}$\xspace}
\newcommand{\atla}{$\textit{ATLAS}$\xspace}
\newcommand{\edit}{$\textsc{EditAS}$\xspace}

\usepackage{orcidlink}

\usepackage{subfigure}
\usepackage{tablefootnote}
\usepackage{colortbl}
\usepackage{balance}
\usepackage[most]{tcolorbox}

\usepackage[normalem]{ulem}
\newcommand{\revise}[1]{{\color{black}{#1}}}
\newcommand{\delete}[1]{}

\newcommand{\newrevise}[1]{{\color{black}{#1}}}
\newcommand{\newdelete}[1]{}

\newcommand{\finding}[2]{
\begin{center}
\begin{tcolorbox}[leftrule=0mm,toprule=0mm,bottomrule=0mm,rightrule=0mm,left=1pt,right=2pt,top=0pt,bottom=0pt,breakable]
\textbf{Answer to RQ{#1}:}
{#2}
\end{tcolorbox}
\end{center}
}

\newcommand{\find}[2]{
\begin{tcolorbox}[toprule=0mm,bottomrule=0mm,left=1pt,right=2pt,top=2pt,bottom=2pt,breakable]%
\em #1
\end{tcolorbox}
}

\AtBeginDocument{%
  \providecommand\BibTeX{{%
    \normalfont B\kern-0.5em{\scshape i\kern-0.25em b}\kern-0.8em\TeX}}}
    
\begin{document}

\title{Improving Retrieval-Augmented Deep Assertion Generation via Joint Training}

\author{Quanjun Zhang
\orcidlink{0000-0002-2495-3805},
Chunrong Fang\orcidlink{0000-0002-9930-7111},
Yi Zheng\orcidlink{0009-0009-2040-8496}, 
Ruixiang Qian\orcidlink{0009-0003-5040-3123}, 
Shengcheng Yu\orcidlink{0000-0003-4640-8637},
Yuan Zhao\orcidlink{0000-0003-1980-6277},
Jianyi Zhou\orcidlink{0000-0002-4867-5416},
Yun Yang\orcidlink{0000-0002-7868-5471},
Tao Zheng\orcidlink{0009-0001-3736-4604}, 
Zhenyu Chen\orcidlink{0000-0002-9592-7022},

\IEEEcompsocitemizethanks{
\IEEEcompsocthanksitem 
Quanjun Zhang and Chunrong Fang
are with the State Key Laboratory for Novel Software Technology, Nanjing University, China, and Shanghai Key Laboratory of Computer Software Evaluating and Testing, Shanghai, China. \protect\\
E-mail: 
quanjun.zhang@smail.nju.edu.cn,
fangchunrong@nju.edu.cn

\IEEEcompsocthanksitem 
Yi Zheng, Ruixiang Qian, ShengCheng Yu, Yuan Zhao, Tao Zheng are with the State Key Laboratory for Novel Software Technology, Nanjing University, China. \protect\\
E-mail: 
201250182@smail.nju.edu.cn,
qianrx@smail.nju.edu.cn,
yusc@gmail.com,
allenzcrazy@gmail.com,
zt@nju.edu.cn,

\IEEEcompsocthanksitem Jianyi Zhou is with Huawei Cloud Computing Technologies Co., Ltd.\protect\\
E-mail: zhoujianyi2@huawei.com.

\IEEEcompsocthanksitem Yun Yang is with the Department of Computing Technologies, Swinburne University of Technology, Melbourne, VIC 3122, Australia.\protect\\
E-mail: yyang@swin.edu.au

Zhenyu Chen
are with the State Key Laboratory for Novel Software Technology, Nanjing University, China and Shenzhen Research Institute of Nanjing University, China \protect\\
E-mail: zychen@nju.edu.cn

\IEEEcompsocthanksitem Chunrong Fang is the corresponding author.

}

\thanks{Manuscript received xxx xxx, 2024; revised xxx xxx, 2025.}
}

\markboth{IEEE Transactions on Software Engineering}%
{Shell \MakeLowercase{\textit{et al.}}: Bare Demo of IEEEtran.cls for Computer Society Journals}

\IEEEtitleabstractindextext{
\begin{abstract}
\justifying
Unit testing attempts to validate the correctness of basic units of the software system under test and has a crucial role in software development and testing.
However, testing experts have to spend a huge amount of effort to write unit test cases manually.
Very recent work proposes a retrieve-and-edit approach to automatically generate unit test oracles, \ie assertions.
Despite being promising, it is still far from perfect due to some limitations, such as splitting assertion retrieval and generation into two separate components without benefiting each other.
In this paper, we propose \toolname{}, a retrieval-augmented automated assertion generation (AG) approach that leverages external codebases and joint training to address various technical limitations of prior work.
Inspired by the plastic surgery hypothesis, \toolname{} attempts to combine relevant unit tests and advanced pre-trained language models (PLMs) with retrieval-augmented fine-tuning.
\revise{
The key insight of AG-RAG is to simultaneously optimize the retriever and the generator as a whole pipeline with a joint training strategy, enabling them to learn from each other.}
Particularly, \toolname{} builds a dense retriever to search for relevant test-assert pairs (TAPs) with semantic matching and a retrieval-augmented generator to synthesize accurate assertions with the focal-test and retrieved TAPs as input.
Besides, \toolname{} leverages a code-aware language model CodeT5 as the cornerstone to facilitate both assertion retrieval and generation tasks.
\delete{
Furthermore, the retriever is optimized in conjunction with the generator as a whole pipeline with a joint training strategy.
This unified design fully adapts both components specifically for retrieving more useful TAPs, thereby generating accurate assertions.}
\revise{Furthermore, \toolname{} designs a joint training strategy that allows the retriever to learn from the feedback provided by the generator.
This unified design fully adapts both components specifically for retrieving more useful TAPs, thereby generating accurate assertions.
}
\toolname{} is a generic framework that can be adapted to various off-the-shelf PLMs.
We extensively evaluate \toolname{} against six state-of-the-art AG approaches on two benchmarks and three metrics.
Experimental results show that \toolname{} significantly outperforms previous AG approaches on all benchmarks and metrics, \eg improving the most recent baseline \edit{} by 20.82\% and 26.98\% in terms of accuracy.
\toolname{} also correctly generates 1739 and 2866 unique assertions that all baselines fail to generate, 3.45X and 9.20X more than \edit{}.
\delete{We further demonstrate the positive contribution of our joint training strategy, \eg improving a variant without the retriever by an average accuracy of 14.11\%, and the generalizability of our framework, \eg \toolname{} with four different PLMs improving \edit{} by an average accuracy of 9.02\%.}
\revise{We further demonstrate the positive contribution of our joint training strategy, \eg \toolname{} improving a variant without the retriever by an average accuracy of 14.11\%.
Besides, adopting other PLMs can provide substantial advancement, \eg \toolname{} with four different PLMs improving \edit{} by an average accuracy of 9.02\%, highlighting the generalizability of our framework.
Overall, our work demonstrates the promising potential of jointly fine-tuning the PLM-based retriever and generator to predict accurate assertions by incorporating external knowledge sources, thereby reducing the manual efforts of unit testing experts in practical scenarios.
}
\end{abstract}

}

\maketitle
\IEEEdisplaynontitleabstractindextext

\IEEEpeerreviewmaketitle

\IEEEraisesectionheading{
\section{Introduction}
\label{sec_intro}
}
\IEEEPARstart{U}nit testing attempts to validate the correctness of software systems by basic functional components or units, which serves as the cornerstone in improving software quality and reliability~\cite{dinella2022toga,fraser2010mutation,shang2024large}.
This practice typically involves writing unit tests to ensure that individual components (\eg methods, classes, and modules) are implemented correctly as designed by developers.
Unlike integration and system testing~\cite{arcuri2019restful}, which assess the entire software system as a whole, including different components and external dependencies, unit testing involves each individual component~\cite{elbaum2002test,schafer2023empirical}.
Thus, unit testing enables the early detection and diagnosis of failures, facilitating a more efficient software development process~\cite{daka2014survey,fraser2014large}.

However, it is fundamentally challenging and labor-intensive for developers to construct high-quality unit tests manually~\cite{daka2014survey}.
To mitigate manual efforts in writing unit tests, a mass of approaches have been proposed to automate test generation~\cite{pacheco2007randoop,fraser2011evosuite,fraser2012seed,fraser2012whole} \revise{such as heuristic-based~\cite{fraser2011evosuite}, random-based~\cite{pacheco2007randoop}, and symbolic execution~\cite{braione2016jbse}}.
A unit test is typically composed of a test prefix, \ie a sequence of statements to invoke the specific behavior of the unit under test, and a test oracle, \ie an assertion statement to specify the expected behavior.
\delete{Despite being promising in generating high-coverage test prefixes, these tools struggle to accurately capture the intended program behavior with meaningful assertions.}
\revise{Despite being promising in generating high-coverage test prefixes, these tools struggle to accurately capture the intended program behavior with meaningful assertions due to the reliance on heuristic or random algorithms.}
\delete{For example, prior work~\cite{almasi2017industrial} reveals that tool-generated assertions are not as meaningful and useful as human-written ones in the industrial scenario.}
\revise{For example, prior work~\cite{almasi2017industrial} reveals that assertions generated by traditional tools (such as EvoSuite~\cite{fraser2011evosuite} and Randoop~\cite{pacheco2007randoop}) are not as meaningful and useful as human-written ones in the industrial scenario.}

To address the crucial unit assertion issue, Watson~\etal~\cite{watson2020learning} introduce \atla{}, the first \textbf{Deep Learning (DL)}-based \textbf{Assertion Generation (AG)} approach, which trains a sequence-to-sequence model with corpora of historical unit tests.
\atla{} takes a focal method (\ie a method under test) and its test prefix as inputs, and returns an assertion as output.
For convenience, we denote the input as a focal-test, and the input-output pair as a \textbf{Test-Assert Pair (TAP)}.
However, \atla{} faces difficulties in generating assertions with low-frequency tokens or a long code sequence.
Furthermore, given a focal-test, Yu~\etal~\cite{yu2022automated} propose a family of retrieval-based AG techniques, namely \irar{}, \rann{}, and \rah{}, to retrieve the most relevant focal-test and its assertion, and produce the final assertion with various adaptation strategies.
They also combine \atla{} and above retrieval-based techniques to propose an integrated AG approach, abbreviated as \inte{}.
However, such retrieval-based approaches may struggle to understand the semantic differences between the given and retrieved focal-tests, leading to inappropriate modifications of the retrieved assertions.

Very recently, Sun~et al.~\cite{sun2023revisiting} propose \edit{}, a retrieve-and-edit approach for automated assertion generation to address various challenges of prior work~\cite{watson2020learning,yu2022automated}.
\edit{} first retrieves a similar focal-test from an external corpus and utilizes a neural sequence-to-sequence model to learn the assertion edit patterns.
Despite being the most competitive AG technique, \edit{} fails to tackle more fundamental challenges of prior work, rendering it still imperfect.
\begin{itemize}[leftmargin=*]
    \item[\ding{172}]\textbf{Assertion Retriever}. \edit{} leverages a sparse term-based retriever to search for relevant assertions based on lexical matching, which is sensitive to the choice of identifier naming in unit tests while failing to consider the meaningful code semantics.

    \item[\ding{173}]\textbf{Assertion Generator}.
    \edit{} trains the generator with a limited code corpus, \eg only 156,760 samples in \olddata{}, which may generate sub-optimal vector representations for unit tests.
    
    \item[\ding{174}]\textbf{Training Paradigm}.
    \edit{} treats the retriever and generator as independent components to either retrieve or generate assertions.
    The pipeline fails to optimize them as a whole pipeline, thus potentially limiting the overall generation performance.
\end{itemize}

In this paper, we propose a novel AG approach called {\toolname} to address the aforementioned limitations of \edit{}.
Our work is motivated by the potential of integrating the well-known plastic surgery hypothesis~\cite{barr2014plastic} with the recent \textbf{Pre-trained Language Models (PLMs)}~\cite{wang2021codet5} in the field of assertion generation.
The plastic surgery hypothesis provides profound implications in software engineering~\cite{barr2014plastic}, \ie in real-world development scenarios, developers usually refer to similar code snippets from open-source projects to assist in generating new code snippets.
To this end, we automate the plastic surgery hypothesis by fine-tuning retrieval-augmented PLMs, \ie retrieving similar assertions from external codebases to assist in fine-tuning PLMs for new assertion generation.
\delete{Particularly, given a focal-test, we leverage a neural encoder as the dense assertion retriever to search for relevant TAPs by measuring their semantic similarity.}
\revise{Particularly, given a focal-test, we design a dense assertion retriever to search for relevant TAPs from external codebases by measuring their semantic similarity.
The dense retriever employs neural networks to encode code semantics, thus capturing hidden and intricate relationships between focal-tests and assertions.}
We then build a retrieval-augmented assertion generator to synthesize accurate assertions with external TAPs to guide the generation process. 
We utilize a code-aware language model, CodeT5~\cite{wang2021codet5}, as the foundation mode of \toolname{} to facilitate both assertion retrieval and generation tasks in a unified manner.
CodeT5 is pre-trained from a mass of open-source projects in the wild to contain general knowledge about programming languages, achieving state-of-the-art performance in both code understanding and generation tasks.
Besides, we further jointly optimize the retriever and the generator with a unified training strategy, to fully adapt them as a whole pipeline for better assertion retrieval and generation.
\revise{The joint training loss is calculated by the generation loss of the retrieved TAPs and their retrieval probabilities, aligning high-probability TAPs retrieved by the retriever with low-loss TAPs propagated by the generator. 
Thus, this strategy enables the retriever to learn to select TAPs based on their feedback in guiding the generator to synthesize ground truth, while guiding the generator to pay more attention to more helpful TAPs based on their retrieval probabilities.}
\toolname{} is generic in concept and can be easily integrated with various encoder-decoder PLMs.
\delete{Although the retrieval-augmented generation pipeline has been explored in prior work~\cite{wang2023rap,nashid2023retrieval,parvez2021retrieval}, we are the first to investigate its effectiveness for assertion generation by using external knowledge sources to jointly fine-tune the PLM-based retriever and generator.}

\delete{
The distinctions between \toolname{} and previous AG approaches~\cite{watson2020learning,yu2022automated,sun2023revisiting} mainly lie in the retriever, the generator, and the training paradigm.
First, prior work utilizes a sparse retriever (\eg Jaccard similarity~\cite{sun2023revisiting}) based on lexical matching, while \toolname{} builds a dense retriever to search for relevant TAPs with more meaningful code semantics.
Besides, prior work trains an assertion generator with \delete{a vanilla transformer}\revise{a basic encoder-decoder model} (\eg RNNs~\cite{watson2020learning}) from a limited number of labeled data, 
while \toolname{} is built upon off-the-shelf CodeT5, which is optimized from a large codebase to obtain meaningful vector representations for unit tests.
Finally, prior work restricts the retriever only to provide similar assertions without benefiting from training, \toolname{} trains the dense retriever to learn how to better guide the generation process with a unified training strategy.
}

We conduct extensive experiments to compare {\toolname} with six state-of-the-art AG approaches (including both DL-based and retrieval-based ones) on two widely adopted benchmarks and three evaluation metrics.
The experimental results demonstrate that {\toolname} is able to outperform all existing AG approaches with an accuracy of 64.59\% and 56.33\%, improving the most recent baseline \edit{} by 20.82\% and 26.98\%.
Besides, {\toolname} successfully generates 1739 and 2866 unique assertions that no prior work can produce, which are 1348 (3.45X) and 2585 (9.20X) more than \edit{}, demonstrating that {\toolname} can complement existing AG approaches well.
Moreover, we implement {\toolname} with three other PLMs (\eg UniXcoder), and find an average of 55.56\% of assertions are correctly generated, highlighting the generalizability of {\toolname}.

To sum up, the contributions of this paper are as follows:
\begin{itemize}
    \item 
    We introduce a generation pipeline for unit assertions, leveraging PLMs through a retrieval-augmented process followed by joint fine-tuning.
    To the best of our knowledge, this is the first work to explore the power of unit test retrieval in external codebases for PLM-based AG approaches.
    
    \item
    We implement \toolname{}, a novel retrieval-augmented PLM-based assertion generation approach.
    \toolname{} utilizes a dense retriever to search for assertions of similar focal-tests as prototypes and employs a generator to learn correct assertions with augmented inputs.
    Both components are jointly optimized with a training strategy as a whole pipeline.
    Importantly, \toolname{} is a generic AG framework and can be integrated with various encoder-decoder PLMs.

    \item
    We conduct extensive experiments with six baselines, two benchmarks, and three metrics to demonstrate that \toolname{} significantly outperforms existing AG approaches.

    \item 
    \newrevise{To facilitate follow-up studies, we open-source a replication package, including datasets, source code, and models~\cite{myurl}.}

\end{itemize}

\section{Background and Related Work}
\label{sec_background}

\subsection{Unit Test Generation}
\revise{
In the literature, to reduce the manual effort involved in writing unit tests, researchers have proposed numerous approaches for automatically generating test cases, including heuristic-based~\cite{fraser2011evosuite}, random-based~\cite{pacheco2007randoop}, and symbolic execution~\cite{braione2016jbse}
Among them, EvoSuite~\cite{fraser2011evosuite} and Randoop~\cite{pacheco2007randoop} are widely regarded as foundational works in the field of automated test generation, providing critical guidance for the conception and development of subsequent approaches.
However, such traditional test generation tools often rely on heuristics or randomness to generate assertion statements without considering the code semantics of focal methods, and thus are limited in their ability to generate useful and meaningful assertions~\cite{shamshiri2015automated, almasi2017industrial}.
For example, Almasi~\etal~\cite{almasi2017industrial} conduct an investigation of EvoSuite and Randoop in an industrial software system, and find that ``the assertions are meaningful and useful unlike the generated ones''.
As a complement to traditional test generation tools, automated assertion generation attempts to synthesize program assertions based on the functions under test.
This topic has garnered significant attention in recent years~\cite{zhai2020c2s, goffi2016automatic, sun2023revisiting, yu2022automated, watson2020learning}, and is the focus of this paper.
}

\label{sec:background_AG}

\subsection{DL-based Assertion Generation}
\label{sec:background_dl_ag}

\begin{figure}[t]
	\graphicspath{{figs/}}
	\centering
	\includegraphics[width=0.99\linewidth]{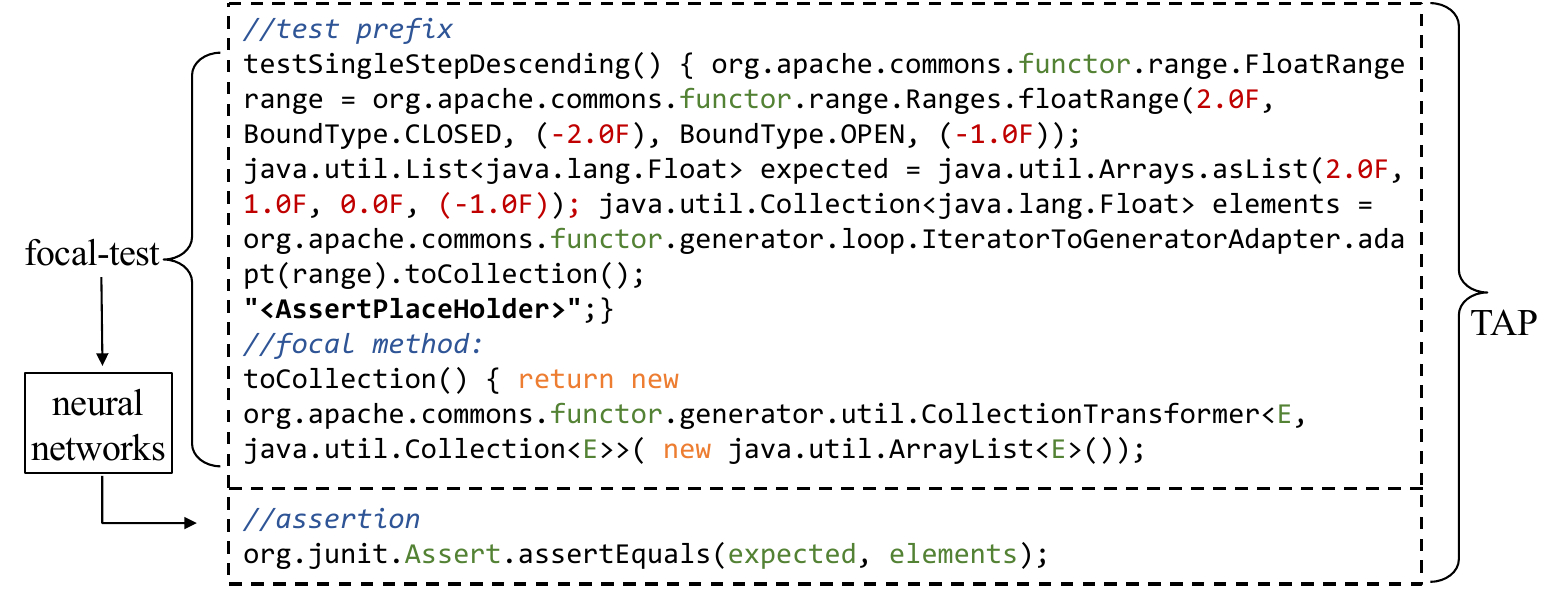}
	\caption{Example of a Test-Assertion Pair (TAP) in \atla{}}
	\label{fig:tap}
\end{figure}

With the success of DL, researchers have increasingly been utilizing advanced DL techniques to automate a variety of software engineering tasks~\cite{watson2022systematic,yang2022survey}, such as program repair~\cite{zhang2023survey,chen2022neural,zhang2023gamma} and vulnerability detection~\cite{cheng2022path,chakraborty2021deep,fu2022vulrepair,cheng2024bug}.
In the era of unit testing, Watson~\etal~\cite{watson2020learning} introduce \atla{}, the first DL-based AG approach to directly predict an accurate assertion from its focal-test by sequence-to-sequence learning.
They first extract a mass of TAPs from open-source projects in the wild and then abstract source code to train a recurrent neural network (RNN) model.
As illustrated in~\figref{fig:tap}, each TAP in \atla{} consists of two components: a focal-test (\ie a focal method denoting the method under test and a test prefix denoting a test method without any oracle) and its assertion.

The community has also seen some preliminary explorations of PLMs in supporting assertion generation~\cite{tufano2022generating, mastropaolo2021studying, mastropaolo2022using,zhang2024exploring}.
For example, Mastropaolo~\etal~\cite{mastropaolo2022using} investigate the performance of the T5 model in supporting four tasks by transfer learning, including bug-fixing, mutant injection, assertion generation, and code summarization.
These studies typically pre-train a language model from scratch with source code or English texts, and fine-tune them to benefit multiple downstream tasks.
On the contrary, we aim to propose a specific AG approach \toolname{} empowered with off-the-shelf PLMs.
Recently, Nashid~\cite{nashid2023retrieval} propose CEDAR, a prompt-based few-shot learning approach for both program repair and assertion generation.
CEDAR queries a large language model (LLM) Codex~\cite{chen2021evaluating} to generate an assertion by constructing a prompt that includes natural language instructions, several examples of task demonstrations, and an output query.
Our work essentially differs from CEDAR regarding (1) the retriever (offline retrieval \textit{v.s.} online optimization); (2) the generator (a black-box billion-level LLM \textit{v.s.} an open-source million-level PLM); and (3) the learning paradigm (few-shot learning with prompt engineering \textit{v.s.} retrieval-augmented fine-tuning). 

\subsection{Retrieval-based Assertion Generation}

Information Retrieval (IR) has been regarded as an effective means to boost the application of DL techniques in software engineering tasks~\cite{parvez2021retrieval,lewis2020retrieval,zhu2022simple,liu2020retrieval}.
Inspired by the integration of IR and DL, Yu~\etal~\cite{yu2022automated} propose a family of retrieval-based approaches for assertion generation:
(1) \irar{} retrieves a TAP with the highest Jaccard similarity on the code token level given a focal-test, and returns its corresponding assertion as output;
(2) $RA_{adapt}$ further replaces incorrect tokens in retrieved assertions from \irar{} with two adaption strategies, \ie a heuristic-based approach \rah{}, and a neural network-based approach \rann{};
and (3) \inte{} builds an inference model to calculate the ``compatibility'' of assertions produced by the above three approaches, and utilizes \atla{} to predict an assertion from scratch if the compatibility is below a pre-defined threshold.
To address limitations of \inte{}, Sun~\etal~\cite{sun2023revisiting} propose \edit{}, a retrieve-and-edit AG approach based on an IR retriever and an LSTM-based sequence-to-sequence model.
\revise{Similar to aforementioned studies~\cite{watson2020learning,yu2022automated,sun2023revisiting}, \toolname{} also follows a \textit{retrieval-and-generation} pipeline; however, the distinction between \toolname{} and prior work mainly lies in three key aspects: the retriever, the generator, and the training paradigm.
First, prior work utilizes a sparse retriever (\eg Jaccard similarity~\cite{sun2023revisiting}) based on lexical matching, while \toolname{} builds a dense retriever to search for relevant TAPs with more meaningful code semantics.
Second, prior work trains an assertion generator with a basic encoder-decoder model (\eg RNNs~\cite{watson2020learning}) from a limited number of labeled data, 
while \toolname{} is built upon off-the-shelf CodeT5, which is optimized from a large codebase to obtain meaningful vector representations for unit tests.
Third, prior work restricts the retriever only to provide similar assertions without benefiting from training, \toolname{} trains the dense retriever to learn how to better guide the generation process with a unified joint
training strategy.}

\delete{Unlike prior work~\cite{yu2022automated,sun2023revisiting} relying on a token-based retriever, \toolname{} utilizes a dense retriever to consider code semantics, which is a parametric model and can be further optimized with the generator.}

\subsection{Pre-Trained Language Model}
\label{sec_background_LLM}

PLMs have demonstrated their potential capabilities to revolutionize a mass of software engineering tasks~\cite{wang2024software,fan2023large,tufano2022using, li2022automating,zhang2024systematic,zhang2023survey2}.
Existing PLMs are fundamentally built with the Transformer architecture~\cite{vaswani2017attention} and are categorized into three main types.
(1) \textbf{Encoder-only PLMs}~\cite{feng2020codebert,guo2020graphcodebert} train the encoder to convert an input into a fixed-size context vector with masked language modeling.
(2) \textbf{Decoder-only PLMs}~\cite{lu2021codexglue} train the decoder to predict the next word in a sequence given the previous word with unidirectional language modeling.
(3) \textbf{Encoder-Decoder PLMs}~\cite{wang2021codet5} train both the encoder and decoder to encode the input sequence and generate an output sequence with denoising objectives.
Overall, encoder-only PLMs are trained to produce bidirectional representations, thus suitable for code understanding, such as vulnerability detection~\cite{fu2022linevul}, while decoder-only LLMs are typically used for auto-regressive generation, such as code completion~\cite{ribeiro2022framing}.
Encoder-decoder LLMs combine the respective advantages of both the encoder and decoder to understand inputs and generate relevant outputs and suitable sequence-to-sequence generation, such as program repair~\cite{wang2023rap,yuan2022circle}.

In this work, we select encoder-decoder PLMs as the foundation model of \toolname{}, because \toolname{} generates assertions in a sequence-to-sequence learning manner.
Following previous PLM-based studies~\cite{wang2023rap,peng2024domain,fu2022vulrepair,zhang2023pre}, we implement \toolname{} with CodeT5, a generic code-aware language model that is pre-trained on a large code corpus, and achieve state-of-the-art performance in both code understanding and generation tasks.
In the literature~\cite{wang2024software}, CodeT5 is the most popular and widely-adopted language model that is fine-tuned to support sequence-to-sequence code generation tasks.

\subsection{\revise{Information Retrieval for SE}}
\label{sec:background_IR}

\revise{
Information Retrieval (IR) involves searching for relevant data within large datasets, typically in response to a specific query. 
IR techniques have been widely applied in various code-related tasks, such as fault localization~\cite{dao2017does} and test case prioritization~\cite{peng2020empirically}.
These techniques aim to identify the most relevant objects within a database by leveraging different similarity measures, such as Jaccard similarity, which quantifies the overlap between elements in two sets.
In recent years, the advent of PLMs has spurred the development of retrieval-augmented paradigms for generation tasks. 
This paradigm has found applications in areas such as program repair~\cite{wang2023rap, nashid2023retrieval, zhang2024no} and code summarization~\cite{li2021editsum, parvez2021retrieval}. 
By incorporating external knowledge through retrieval, this paradigm enhances the quality of generated outputs, supplementing the internal knowledge representation of PLMs~\cite{lewis2020retrieval}.
Although the retrieval-augmented generation pipeline has been explored in prior work~\cite{wang2023rap,nashid2023retrieval,parvez2021retrieval}, we are the first to investigate its effectiveness for assertion generation by using external knowledge sources to fine-tune the PLM-based retriever and generator jointly.
}

\section{Approach}
\label{sec_approach}

\begin{figure*}
    \centering
    \includegraphics[width=0.8\linewidth]{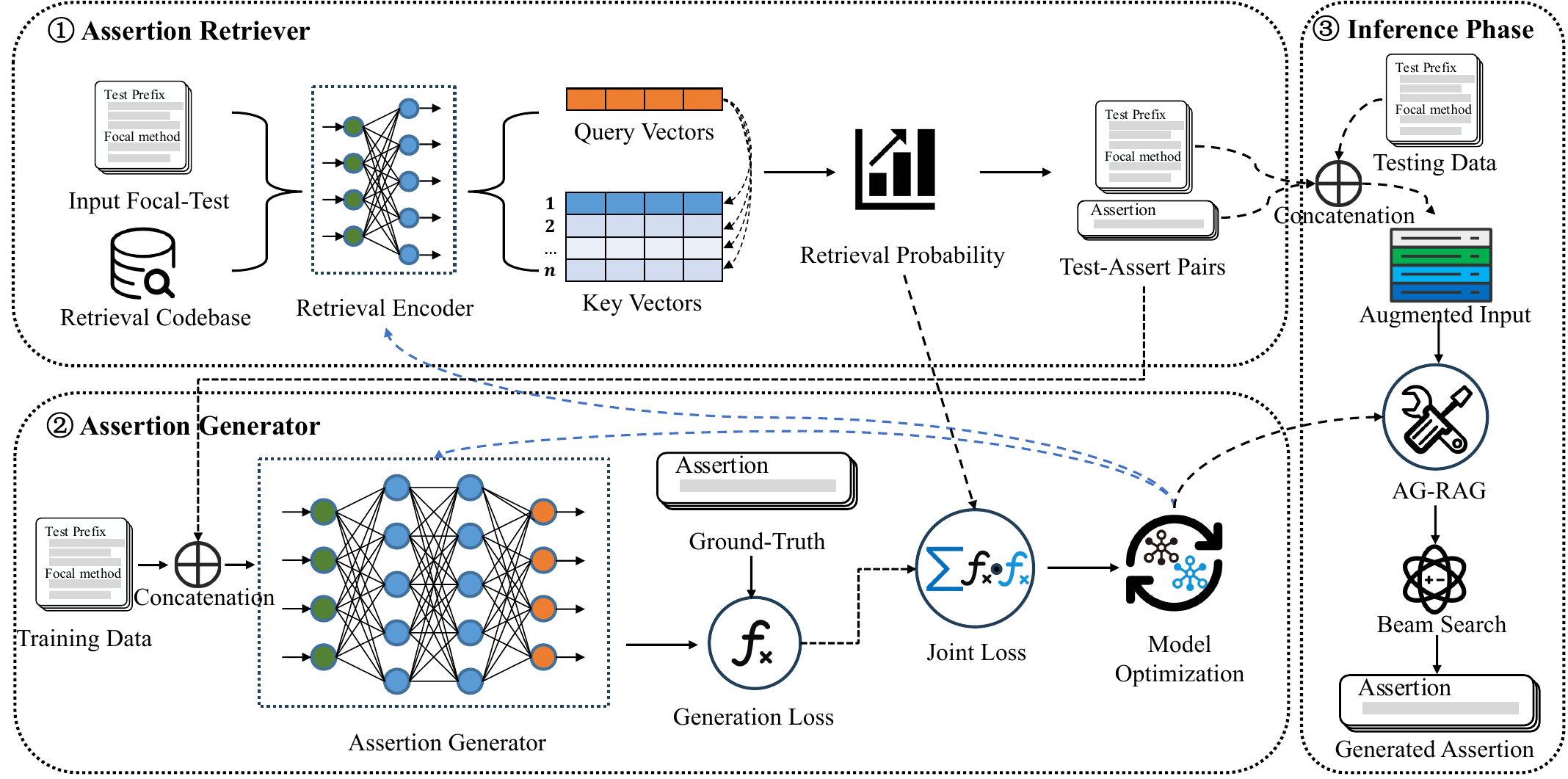}
    \caption{The overall framework of \toolname{}}
    \label{fig:workflow}
\end{figure*}

The overall framework of \toolname{} is illustrated in \figref{fig:workflow}.
In the assertion retrieval stage, 
\toolname{} searches for similar TAPs from the external codebase by calculating their semantic similarity by a dense retriever.
In the assertion generation stage, 
\toolname{} fine-tunes a pre-trained encoder-decoder generator with the retrieval-augmented inputs, \ie the original focal-test and similar TAPs.
During the training process, both the retriever and generator are optimized with a novel joint training strategy, which better adapts them as a whole pipeline to our task.
In the assertion inference stage, after the retriever and a generator are well trained, given a focal-test input and retrieval codebase, the beam search strategy is leveraged to generate a ranked list of candidate assertions and return the one with the highest probability of being correct.

\subsection{Task Formulation}

Similar to the pioneering work \atla{} in the DL-based AG field, \toolname{} treats the assertion generation task as a sequence-to-sequence problem with an encoder-decoder Transformer, which takes a focal-test as input and an accurate assertion as output.
Suppose $\mathcal{D} = ({FT_i, A_i)}^{|\mathcal{D}|}_{i=1}$ be a unit testing dataset consisting of $|\mathcal{D}|$ TAPs,
where $FT_i$ and $A_i$ are the $i$-th focal-test and its corresponding assertion.
The assertion generator attempts to predict $A_i$ from $FT_i$ in a sequence-to-sequence manner, formally defined as follows:

\find{
\begin{definition}
\label{def:generation}
\textbf{Deep Assertion Generation:}\\
Given a focal-test input $FT_i=\left[ft_1, \cdots, ft_m\right]$ with $m$ code tokens and an assertion output $A_i=\left[a_1, \ldots, a_n\right]$ with $n$ code tokens, 
the problem of deep assertion generation is formalized to maximize the conditional probability of $A_i$ being the correct assertion:
\begin{displaymath}
\small
P_{\theta}(A_i|FT_i)=\prod_{j=1}^{n}P_{\theta}
(a_j|a_1, \cdots, a_{j-1}; ft_1, \cdots, ft_m) 
\end{displaymath}
\label{def:ag}
\end{definition}
}

However, unlike \atla{} that directly predicts an assertion $A_i$ from the focal-test $FT_i$, \toolname{} augments the input with additional retrieved TAPs to guide the generation process.
Assume that we have an external codebase containing a collection of historical TAPs $\mathcal{C} = {(FT'_j, A'_j)}^{|\mathcal{C}|}_{j=1}$, where $FT'_j$ and $A'_j$ denotes the $j$-th previous focal-test, and its assertion.
Based on Definition~\ref{def:generation}, the retrieval-augmented deep assertion generation can be formulated as follows:

\find{
\begin{definition}\textbf{\small Retrieval-Augmented Deep Assertion Generation:}\\ 
Given a focal-test $FT_i$ in $\mathcal{D}$, the retriever of \toolname{} searches for the most relevant focal-test $FT'_j$ from the codebase $\mathcal{C}$, as well as its assertion $A'_j$.
Then the original focal-test input $FT_i $ is augmented with the retrieved TAP to form a new input sequence $\hat{FT_i}=FT_i \oplus FT'_j \oplus A'_j $, where $\oplus $ denotes the concatenation operation.
Finally, the assertion generator of \toolname{} attempts to generates $A_i$ from $\hat{FT_i}$ by learning the following probability parameterized by $\theta$:
\begin{displaymath}
P_{\theta}(A_i|\hat{FT_i})=\prod_{j=1}^{n}P_{\theta}
(a_j|a_1, \cdots, a_{j-1}; \underbrace{FT_i}_{\text{Original}}; \underbrace{FT'_j; A'_j }_{\text{Augmented}}) 
\end{displaymath}
\end{definition}
}

\subsection{Dense Assertion Retriever}
\label{sec:approach_retrieval}

The retriever module of \toolname{} attempts to retrieve relevant TAPs from external codebases to guide the generation process.
As mentioned in Section~\ref{sec_background}, most previous AG studies use traditional IR techniques (\eg Jaccard similarity) that only consider word overlaps without considering code semantics.
Besides, these sparse retrievers are not parametric models and cannot be optimized during training.
To enable the joint training of both the retriever and the generator, we employ a dense retriever to search for relevant TAPs by measuring their semantic similarity.

\delete{To encode the given focal-test, we use a Transformer-based encoder to map each focal-test to a fixed-size dense vector.}
\revise{To encode the given focal-test, we initialize a Transformer-based encoder to map each focal-test to a fixed-size dense vector, which is initialized from a pre-trained CodeT5 encoder.}
Particularly, \toolname{} leverages a code-specific Byte-Pair-Encoding (BPE) tokenizer to split the source code of the focal-test into a sequence of tokens.
Unlike the default word-level tokenizer, which adds the full tokens directly to the vocabulary, the BPE tokenizer splits tokens into multiple subwords based on their frequency distribution.
As a result, it can reduce the vocabulary size and mitigate the long-standing Out-Of-Vocabulary (OOV) problem in the code domain.
Besides, \toolname{} prepends a special token of [CLS] into its tokenized sequence, and calculates the final layer hidden state of the [CLS] token as the contextual embedding.
We use a shared encoder to separately encode the query focal-test $FT_i$ in $\mathcal{D}$ and a key focal-test $FT'_j$ in $\mathcal{C}$ as $CLS_{FT_i}$ and $CLS_{FT'_j}$, respectively.
Then, \toolname{} leverages L2 normalization to the embeddings to facilitate the convergence of model training.
L2 normalization adjusts the embeddings' magnitudes, encouraging a more uniform distribution of the feature space, as defined in Equation~\ref{equ:l2}.

\begin{equation}
\small
CLS_{FT}^{\prime}=\frac{CLS_{FT}}{\sqrt{\sum_{j=1}^d CLS_{FT}^2}}
\label{equ:l2}
\end{equation}
where $d$ is the hidden size of the encoder.
Finally, {\toolname} measures the semantic relevance of two focal-tests by calculating the inner product between their normalized embeddings (\ie Cosine similarity), as defined in Equation~\ref{equ:seman_simi}.
 
\begin{equation}
\small
f_\phi\left(FT_i, FT_j\right)=
\left[CLS_{FT_i}^{\prime}\right]^T
\left[CLS_{FT'_j}^{\prime}\right]
\label{equ:seman_simi}
\end{equation}

\subsection{Retrieval-Augmented Assertion Generator}
\label{sec:approach_generation}

The generator module of \toolname{} attempts to produce an accurate assertion based on its focal-test and the relevant TAPs returned by the retriever.
As shown in~\figref{fig:workflow}, given a focal-test $FT_i$, we search for a top relevant TAP $(FT_j, A_j)$, and pass it to the assertion generator to generate an accurate assertion $A_i$. 
We adopt a simple yet effective strategy to augment $FT_i$ into $\hat{FT}_i = FT_i \oplus FT_j \oplus A_j$  via appending the retrieved test-assertion pair into the source focal-test.
Different from prior studies that directly adopt a generator optimized from scratch~\cite{watson2020learning,sun2023revisiting,yu2022automated}, we propose to employ CodeT5, a code-aware language model pre-trained with source code as the cornerstone of the generator.

\textbf{Representation.}
\toolname{} constructs the retrieval-augmented input to the CodeT5 assertion generator as Equation~\ref{equ:input}, where $\cdot$ denotes the concatenation operator, and ``\texttt{\textbackslash n}'' is inserted to separate three parts.
To capture the encoding for the retrieved TAPs, we leverage CodeT5's bimodal capability of processing both programming and natural language (comments) inputs.
We format the retrieved content into a comment by inserting a special token \/\/ at the beginning of it. 

\begin{equation}
    \hat{FT}_i = \texttt{[CLS]} \cdot X_i \cdot \backslash n \cdot // \cdot FT_j \cdot \backslash n \cdot A_j
\label{equ:input}
\end{equation}

\textbf{Model Architecture.}
We build the generation model of \toolname{} with an encoder-decoder Transformer architecture, which consists of an encoder stack, a decoder stack, and a linear layer with a softmax activation function.
The encoder takes $\hat{FT}_i$ as input and emits the accurate assertion $A_i$ from its decoder in an autoregressive manner.
Particularly, \toolname{} splits the source code of the input $\hat{FT_i}$ into subwords using a code-specific BPE tokenizer, as mentioned in Section~\ref{sec:approach_retrieval}.
The utilized tokenizer is pre-trained with eight popular programming languages from CodeSearchNet, making it suitable for tokenizing the source code of TAPs~\cite{wang2021codet5}.
Besides, \toolname{} performs word embedding to obtain contextual representations for tokenized tokens, which are then processed by an encoder stack to derive the hidden state.
Furthermore, the decoder stack's output is directed through a linear layer equipped with a softmax activation function to calculate the likelihood of each token being the next part of the code sequence.

\textbf{Generation Loss.}
We train the assertion generator to learn the transformation rules from the retrieval-augmented input $\hat{FT_i}$ to the output $A_i$ by the sequence-to-sequence learning.
To this end, we leverage \emph{teacher forcing} to minimize the common cross-entropy loss $\mathcal{L}_{ge}$ across all training samples~\cite{wang2023rap}.
As illustrated in Equation~\ref{equation:loss}, the loss value is calculated by comparing each position in the predicted assertion and each position
in the ground-truth assertion.

\begin{equation}
\small
    \mathcal{L}_{ce} = -\sum_{i=1}^{|\mathcal{D}|}\log(P_{\theta}((A_i|\hat{FT_i}))
\label{equation:loss}
\end{equation}

\subsection{Joint Retriever and Generator Training}
\label{sec:approach_joint}

Sections~\ref{sec:approach_retrieval} and ~\ref{sec:approach_generation} have illustrated the core assertion retrieval and generation modules of \toolname{}.
However, the retriever and generator are built independently, so they may not always identify the TAP that would be most beneficial for the generator to produce an accurate assertion.
To further connect the two core components, we attempt to optimize them jointly using a unified training strategy.
Although the external codebase $C$ is a non-parametric memory from which the retrieved TAPs are responsible for guiding the assertion generation process, we utilize a parametric model (\ie a Transformer encoder) as the dense retriever.
The retrieval can be formulated as a latent variable $P_{\phi}(\mathcal{C}|{FT_i})$ from a probabilistic perspective, where $\mathcal{C}$ contains $|\mathcal{C}|$ TAPs.
Then, we can decompose the original deep assertion generation probability $P(A_i|FT_i)$ in Definition~\ref{def:ag} into the retrieval probability and the retrieval-augmented probability,
\ie 
the production of the marginal distribution $P_{\phi}(\mathcal{C}|{FT_i})$ that represents the likelihood of retrieved TAPs given the focal test $FT_i$,
and the condition distribution $P_{\phi}(A_i|{FT_i},\mathcal{C})$ that represents the probability of the assertion $A_i$ given both the focal-test $FT_i$ and the retrieved TAPs.
Formally, by assuming that all TAPs in $\mathcal{C}$ are independent, the generation probability $P(A_i|FT_i)$ is defined as follows.
\looseness=-1

\begin{equation}
\begin{aligned}
\small 
  P(A_i|FT_i) &= P_{\phi}(\mathcal{C}|{FT_i}) \cdot P_{\theta}(A_i|{FT_i},\mathcal{C}) \\
  &=\sum_{j=1}^{|\mathcal{C}|} 
  \underbrace{P_{\phi}(TAP_j|FT_i)}_{\text{Retriever}} \cdot
  \underbrace{P_{\theta}(A_i|FT_i, TAP_j)}_{\text{Generator}} 
\end{aligned}
\end{equation}

\noindent where $P_{\phi}(TAP_j|FT_i)$ denotes the retrieval probability of the $j$-th TAP returned by the retriever $P_{\phi}(\mathcal{C}|{FT_i})$.
To calculate $P_{\phi}(TAP_j|FT_i)$, we use a softmax function to convert their semantic relevance into a probability distribution, defined as follows.

\begin{equation}
\small
P_{\phi}(TAP_j|FT_i)=\frac{f_\phi\left(FT_i, FT_j\right)}{\sum_{k=1}^{|\mathcal{C}|} f_\phi\left(FT_i, FT_k\right)}
\end{equation}
where $f_\phi\left(FT_i, FT_j\right)$ denotes the semantic similarity of the key focal-test $FT_j$ in $TAP_j$ and the query focal-test $FT_i$, which is illustrated in Equation~\ref{equ:seman_simi}.
While it is time-consuming and complicated to calculate the above probabilities and optimize parameters by backpropagation by querying each sample in the large external codebase,
To facilitate efficient training, we consider the top-$k$ TAPs with the highest retrieval probabilities to approximate the above $P(A_i|FT_i)$, defined as follows.

\begin{equation}
\small 
  P(A_i|FT_i)  \approx   
  \sum_{j=1}^{k} 
  P_{\phi}(TAP_j|FT_i) \cdot
  P_{\theta}(A_i|FT_i, TAP_j)
\end{equation}

Up to this point, we can achieve the joint training of the retriever and the generator by minimizing the negative log-likelihood of $P(A_i|FT_i)$.
Formally, during the back-propagation process, we define the training loss function of \toolname{} based on the top-$k$ retrieved TAPs as follows.

\begin{equation}
\small
    \mathcal{L} = \sum_{j=1}^{k}\mathcal{L}_{ce} \cdot P_{\phi}(TAP_j|FT_i)
\end{equation}

\subsection{Assertion Inference}

During the inference stage, as illustrated in~\figref{fig:workflow}, we leverage the beam search strategy to synthesize a ranked list of assertion candidates for a given focal-test.
At each decoding timestep, the beam search uses a best-first search strategy to select the most accurate assertion candidates with the highest estimated likelihood scores.
When an [EOS] token representing the end of the output sequence is emitted, the search process is terminated, and the assertion with the highest score is returned.
The correctness of the returned assertion can be validated by automatically comparing it with the ground truth or manually inspecting it against test experts.

\subsection{\revise{Usage of \toolname{}}}

\revise{
As indicated in Fig.~\ref{fig:workflow}, \toolname{} is intended to be used for predicting accurate assertion statements within an environment that provides focal methods and test prefixes.
In practice, \toolname{} can be deployed in two scenarios: manual and automatic scenarios.
First, in the manual generation scenario, when a developer needs to validate the correctness of basic units of the software system under test, they start by writing a test prefix, which is essentially a sequence of call statements to invoke the specific behavior of the unit under test.
Then, because writing assertions that describe the correct behavior of a program requires an in-depth understanding of the program's functionality and specifications, the developer can directly utilize \toolname{} to generate the corresponding test assertions.
For example, it is possible to integrate \toolname{} in an IDE as a test completion plug-in, which can retrieve similar focal-tests from an external codebase and complete missing assertions based on human-written test prefixes, so as to reduce manual efforts.
Second, a more straightforward application is integrating \toolname{} with existing test generation tools, to enhance the usability of such tools.
Existing automated test case generation tools are able to generate effective test prefixes that achieve high coverage based on heuristic algorithms.
However, such tools usually struggle to understand the intended program behavior,  making it challenging to produce meaningful assertions.
\toolname{} can form an excellent complement to existing test generation tools, making it possible to generate high-quality test cases with bug detection capabilities. 
}

\section{Experimental Setup}
\label{sec_experimental setup}

\subsection{Research Questions}
We conduct experiments to answer three research questions (RQs):
\begin{itemize}
    \item RQ1: How does {\toolname} perform compared to state-of-the-art assertion generation approaches?
    \item RQ2: To what extent does the joint training strategy affect the overall effectiveness of {\toolname}?
    \item RQ3: What is the generalizability of {\toolname} when employing other advanced PLMs?
\end{itemize}

\subsection{Datasets}
\label{sec_dataset}
\begin{table*}[htbp]
  \centering
  \caption{Detailed statistics of each assertion type in $Data_{new}$ and $Data_{old}$}
  \resizebox{0.98\linewidth}{!}{
    \begin{tabular}{c|c|ccccccccc}
    \toprule[0.7pt]
    \textbf{AssertType} & Total & Equals & True  & That  & NotNull & False & Null  & ArrayEquals & Same  & Other \\ \hline
    $Data_{old}$ & 15,676 & 7,866 (50\%) & 2,783 (18\%) & 1,441 (9\%) & 1,162 (7\%) & 1,006 (6\%) & 798 (5\%) & 307 (2\%) & 311 (2\%) & 2 (0\%) \\
    $Data_{new}$ & 26,542 & 12,557 (47\%) & 3,652 (14\%) & 3,532 (13\%) & 1,284 (5\%) & 1,071 (4\%) & 735 (3\%) & 362 (1\%) & 319 (1\%) & 3,030 (11\%) \\
    \bottomrule[0.7pt]
    \end{tabular}
    }
  \label{tab:statistics}
\end{table*}

We select two publicly available datasets to evaluate the assertion generation capability of \toolname{} and baselines, \ie \olddata{}~\cite{watson2020learning} \newdata{}~\cite{yu2022automated}. 
The datasets are large-scale, representative in the community~\cite{mastropaolo2021studying,mastropaolo2022using,nashid2023retrieval,tufano2022generating}, and are utilized by all baselines~\cite{watson2020learning,yu2022automated,sun2023revisiting}.

$\bullet$ \olddata{}~\cite{watson2020learning} is constructed by Watson~\etal~\cite{watson2020learning} and is the first benchmark to evaluate deep assertion approaches.
Watson~\etal~\cite{watson2020learning} first mine more than 9K open-source projects, and extract 2.5 million developer-written test methods.
They then filter out test methods with more than 1K tokens and with unknown tokens.

$\bullet$ \newdata{}~\cite{yu2022automated} is an extended dataset of \olddata{} by including cases that are excluded due to unknown tokens.
\olddata{} excludes assertions containing unknown tokens to oversimplify the assertion generation problem, thus being unsuitable to reflect the real-world data distribution.
To address this issue, Yu~\etal~\cite{yu2022automated} conduct \newdata{} by incorporating an additional 108,660 samples with unknown tokens into the existing samples in \olddata{}.

\revise{
Overall, \olddata{} and \newdata{} contain a total of 156,760 and 265,420 samples, respectively. 
These datasets are divided into training, validation, and test sets using an 8:1:1 ratio, as done by Watson~\etal~\cite{watson2020learning} and Yu~\etal~\cite{yu2022automated}. 
In this paper, we strictly adhere to the replication package provided by prior work~\cite{watson2020learning,yu2022automated, sun2023revisiting} to ensure a fair comparison.
}
The statistics of the test sets for the two datasets are presented in Table~\ref{tab:statistics}, including their distribution across different types.

\subsection{Baselines}
\label{sec:baselines}

To address the above-mentioned RQs, we compare \toolname{} with six state-of-the-art AG approaches from different categories.
We first select \atla{}~\cite{watson2020learning}, the first DL-based AG technique that predicts assertions from input focal-tests directly with sequence-to-sequence learning.
We also consider three retrieval-based AG techniques~\cite{yu2022automated}: \irar{}, \rah{}, and \rann{} that leverage external code bases to retrieve similar assertions.
Finally, we consider one integrated AG approach \inte{}~\cite{yu2022automated}and the most recent follow-up \edit{}~\cite{sun2023revisiting}.
It is worth noting that we exclude CEDAR~\cite{nashid2023retrieval} as a baseline primarily due to the data leakage and reproducibility issues of Codex.
Black-box LLMs, like Codex, are close-sourced with unknown training details, \eg pre-training corpora, and have been proven to suffer from data leakage in code-related tasks~\cite{tian2024debugbench,silva2024gitbug,zhang2023critical}.
Besides, CEDAR fails to release generated assertions~\cite{cedar}, and we cannot replicate the results due to the updates to Codex in OpenAI's API~\cite{codex}, posing a challenge to conduct comparisons.
Thus, we directly follow the experimental design of the most recent AG approach \edit{}.

\subsection{Evaluation Metrics}
\label{sec_metrics}

We consider three metrics to evaluate the correctness and quality of generated assertions, \ie accuracy, BLEU and CodeBLEU.
The first two are the same as those used in previous studies~\cite{watson2020learning,yu2022automated,sun2023revisiting}, while the third one is included additionally in our experiment.

$\bullet$ Accuracy is defined as the proportion of the samples correctly predicted by \toolname{} and baselines among the number of total testing samples.
A generated assertion is considered to be correct if each position of it exactly matches that of the ground truth.

$\bullet$ BLEU measures the syntax similarity between the predicted assertion and the ground truth.
It is calculated by the modified $n$-gram precision of a generated sequence to the reference.
sequence. 

$\bullet$ CodeBLEU~\cite{lu2021codexglue} denotes a code-aware variant of BLEU, specifically tailored for evaluating the quality of auto-generated code~\cite{lu2021codexglue}.
Unlike BLEU, CodeBLEU further incorporates syntactic similarity via AST information and semantic similarity via data-flow analysis, making it more suitable for the AG task.

\subsection{Implementation Details}

To implement \toolname{}, we leverage CodeT5-base with 220M parameters to initialize the generator and its encoder to initialize the retriever.
The hidden dimension is 768, the number of encoder layers is 12, and the number of decoder layers is 12 according to CodeT5~\cite{wang2021codet5}.
The number of retrieved TAPs is set to 5 by default for each focal-test considering time and resource constraints.
We implement \toolname{} with PyTorch~\cite{PyTorch} and perform training with Adam Optimizer.
We set the batch size to 8, the maximum lengths of input to 512, the maximum lengths of output to 64, and the learning rate to 5e-5, all of which are default parameters in CodeT5.
We train \toolname{} for up to 20 epochs and will stop the training process early if the BLEU score on the validation set does not increase within three consecutive epochs.
During inference, we employ a beam search with a beam size of 10 and return the top-1 assertion as the final result.
We conduct all experiments with two NVIDIA GeForce RTX 4090 GPUs on one Ubuntu 20.04 server.
\delete{For consistency with prior work~\cite{yu2022automated,sun2023revisiting}, we use the training set as the retrieval corpus.}
\revise{To ensure that \toolname{} and all baselines are evaluated with the same experimental setup, following existing studies~\cite{yu2022automated,sun2023revisiting,watson2020learning}, we train \toolname{} by the training sets of the two datasets separately, and evaluate \toolname{} with their respective test sets.
Besides, during the retrieval process, for consistency with \edit{}~\cite{sun2023revisiting} and \inte{}~\cite{yu2022automated}, we use the training sets of the two datasets as the retrieval corpus.
}

\begin{table*}[t]
  \centering
  \caption{Comparisons of \toolname{} with state-of-the-art AG approaches}
  \resizebox{0.9\linewidth}{!}{
    \begin{tabular}{c|ccc|ccc}
    \toprule
    \multirow{2}[3]{*}{\textbf{Appraoch}} & \multicolumn{3}{c|}{\olddata{}} & \multicolumn{3}{c}{\newdata{}} \\
\cmidrule{2-7}          
    & \textbf{Accuracy} & \textbf{CodeBLEU} & \textbf{BLEU} & \textbf{Accuracy} & \textbf{CodeBLEU} & \textbf{BLEU} \\
    \midrule
    \atla{} & 31.42\% ($\uparrow$105.57\%) & 63.60\% ($\uparrow$27.14\%) & 68.51\% ($\uparrow$23.76\%) & 21.66\% ($\uparrow$160.06\%) & 37.91\% ($\uparrow$79.66\%) & 37.91\% ($\uparrow$92.19\%) \\
    \irar{} & 36.26\% ($\uparrow$78.13\%) & 71.03\% ($\uparrow$13.84\%) & 71.49\% ($\uparrow$18.60\%) & 37.90\% ($\uparrow$48.63\%) & 62.67\% ($\uparrow$8.68\%) & 57.98\% ($\uparrow$25.66\%) \\
    \rah{} & 40.97\% ($\uparrow$57.65\%) & 72.46\% ($\uparrow$11.59\%) & 73.28\% ($\uparrow$15.71\%) & 39.65\% ($\uparrow$42.07\%) & 63.66\% ($\uparrow$6.99\%) & 59.81\% ($\uparrow$21.82\%) \\
    \rann{} & 43.63\% ($\uparrow$59.36\%) & 72.12\% ($\uparrow$12.12\%) & 73.95\% ($\uparrow$14.66\%) & 40.53\% ($\uparrow$29.11\%) & 63.19\% ($\uparrow$7.79\%) & 59.81\% ($\uparrow$21.82\%) \\
    \inte{} & 46.54\% ($\uparrow$38.78\%) & 73.29\% ($\uparrow$10.33\%) & 78.86\% ($\uparrow$7.52\%) & 42.20\% ($\uparrow$33.48\%) & 63.00\% ($\uparrow$8.11\%) & 60.92\% ($\uparrow$19.60\%) \\
    \edit{} & 53.46\% ($\uparrow$20.82\%) & 77.00\% ($\uparrow$5.01\%) & 80.77\% ($\uparrow$4.98\%) & 44.36\% ($\uparrow$26.98\%) & 64.40\% ($\uparrow$5.76\%) & 63.46\% ($\uparrow$14.81\%) \\
    \midrule
    \toolname{} & \textbf{64.59\%} & \textbf{80.86\%} & \textbf{84.79\%} & \textbf{56.33\%} & \textbf{68.11\%} & \textbf{72.86\%} \\
    \bottomrule
    \multicolumn{5}{l}{$\uparrow$ denotes performance improvement of \toolname{} against state-of-the-art baselines}
    \end{tabular}%
    }
  \label{tab:rq1}
\end{table*}%

\section{Evaluation and Results}
\label{sec_results}

\subsection{RQ1:~Comparison with State-of-the-arts}
\textbf{\emph{Experimental Design.}} 
In RQ1, we aim to evaluate the effectiveness of assertions generated by {\toolname}.
We include \toolname{} against six prior AG approaches , \ie \atla{}, \irar{}, \rah{}, \rann{}, \inte{} and \edit{}, two benchmarks, \ie \olddata{} and \newdata{}, and three evaluation metrics, \ie accuracy, BLEU and CodeBLEU.

\textbf{\emph{Results.}}
Table~\ref{tab:rq1} presents the comparison results of \toolname{} and baselines across three metrics.
Overall, we find that \toolname{} achieves an accuracy of 56.33\%-64.59\%, a CodeBLEU score of 68.11\%-80.86\%, and a BLEU score of 72.86\%-84.79\% on two benchmarks, outperforming all baselines by 20.82\%-160.06\%, 5.01\%-79.66\%, and 4.98\%-92.19\%, respectively.
First, when compared with the DL-based approach \atla{}, \toolname{} yields an improvement of 105.57\%, 27.14\% and 23.76\% for accuracy, \delete{CodeBlEU}\revise{CodeBLEU} and BLEU on \olddata{}.
\revise{The possible reason is that, despite both \toolname{} and \atla{} treating assertion generation as a sequence-to-sequence task, \atla{} needs to generate an assertion from scratch with \delete{a Vanilla Transformer}\revise{a basic encoder-decoder model}, while \toolname{} benefits from a retrieved assertion and a pre-trained encoder-decoder architecture.}
Second, when compared with the retrieval-based approaches, \toolname{} outperforms \irar{}, \rah{}, and \rann{} by 32.26\%, 25.97\%, and 24.14\% across three metrics and two benchmarks on average.
We also find that \irar{} that naively regards retrieved assertions as the final outputs achieves better performance than \atla{}, \revise{indicating that it is reasonable for us to reuse the valuable retrieved assertion as the prototype to guide the generation process.}
Third, when compared with integration-based approaches, \toolname{} is superior to \inte{} and \edit{} with an accuracy improvement of 38.78\% and 20.82\% on \olddata{}, 33.48\% and 26.98\% on \newdata{}.
Similar improvements can be observed on other metrics.
\revise{Importantly, different from the closest competitor \edit{}, \toolname{} leverages Codet5 as the foundational skeleton to capture the meaningful semantics of assertions, and the joint training strategy to fully connect the two components, \ie the assertion retriever and generator, thus achieving the best prediction performance.}

\begin{table*}[htbp]
  \centering
  \caption{Detailed statistics of \toolname{} and baselines for each assert type}
    \resizebox{0.9\linewidth}{!}{
    \begin{tabular}{c|c|c|ccccccccc}
    \toprule
    \multirow{2}[4]{*}{\textbf{Dataset}} & \multirow{2}[4]{*}{\textbf{Approach}} & \multirow{2}[4]{*}{\textbf{Total}} & \multicolumn{9}{c}{\textbf{AssertType}} \\
\cmidrule{4-12}          &       &       & \textbf{Equals} & \textbf{TRUE} & \textbf{That} & \textbf{NotNull} & \textbf{FALSE} & \textbf{Null} & \textbf{ArrayEquals} & \textbf{Same} & \textbf{Other} \\
    \midrule
    \multirow{7}[2]{*}{\olddata{}} & \atla{} & 4925(31\%) & 2501(32\%) & 966(35\%) & 248(17\%) & 598(51\%) & 229(23\%) & 236(30\%) & 100(33\%) & 47(15\%) & 0(0\%) \\
          & \irar{} & 5684(36\%) & 2957(38\%) & 1039(37\%) & 449(31\%) & 439(38\%) & 314(31\%) & 285(36\%) & 111(36\%) & 89(29\%) & \textbf{1(50\%)} \\
          & \rah{} & 6423(41\%) & 3300(42\%) & 1151(41\%) & 536(37\%) & 553(48\%) & 335(33\%) & 316(40\%) & 120(39\%) & 111(36\%) & \textbf{1(50\%)} \\
          & \rann{} & 6839(44\%) & 3509(45\%) & 1225(44\%) & 551(38\%) & 610(52\%) & 342(34\%) & 341(43\%) & 134(44\%) & 126(41\%) & \textbf{1(50\%)} \\
          & \inte{} & 7295(47\%) & 3714(47\%) & 1333(48\%) & 546(38\%) & 724(62\%) & 348(35\%) & 352(44\%) & 148(48\%) & 129(41\%) & \textbf{1(50\%)} \\
          & \edit{} & 8380(53\%) & 4131(53\%) & 1581(57\%) & 526(36\%) & 807(69\%) & 577(57\%) & 469(59\%) & 167(54\%) & 122(39\%) & 0(0\%) \\
          & \toolname{} & \textbf{10125(65\%)} & \textbf{4993(63\%)} & \textbf{1790(64\%)} & \textbf{831(58\%)} & \textbf{853(73\%)} & \textbf{691(69\%)} & \textbf{563(71\%)} & \textbf{204(66\%)} & \textbf{199(64\%)} & \textbf{1(50\%)} \\
    \midrule
    \multirow{7}[2]{*}{\newdata{}} & \atla{} & 5749(22\%) & 2900(23\%) & 619(17\%) & 537(15\%) & 388(30\%) & 126(12\%) & 85(12\%) & 47(13\%) & 37(12\%) & 1010(33\%) \\
          & \irar{} & 10059(38\%) & 4664(37\%) & 1436(39\%) & 1070(30\%) & 600(47\%) & 394(37\%) & 286(39\%) & 147(41\%) & 113(35\%) & 1349(45\%) \\
          & \rah{} & 10525(40\%) & 4882(39\%) & 1487(41\%) & 1142(32\%) & 651(51\%) & 403(38\%) & 297(40\%) & 154(43\%) & 121(38\%) & 1388(46\%) \\
          & \rann{} & 10758(41\%) & 4988(40\%) & 1526(42\%) & 1161(33\%) & 691(54\%) & 401(37\%) & 308(42\%) & 162(45\%) & 126(39\%) & 1395(46\%) \\
          & \inte{} & 11201(42\%) & 5248(42\%) & 1566(43\%) & 1196(34\%) & 711(55\%) & 401(37\%) & 313(43\%) & 162(45\%) & 128(40\%) & 1476(49\%) \\
          & \edit{} & 11773(44\%) & 5339(42\%) & 1702(47\%) & 1304(37\%) & 800(62\%) & 523(49\%) & 376(51\%) & 172(47\%) & 139(44\%) & 1418(47\%) \\
          & \toolname{} & \textbf{14950(56\%)} & \textbf{6938(55\%)} & \textbf{2055(56\%)} & \textbf{1832(52\%)} & \textbf{884(69\%)} & \textbf{676(63\%)} & \textbf{447(61\%)} & \textbf{203(56\%)} & \textbf{188(59\%)} & \textbf{1727(58\%)} \\
    \bottomrule
    \end{tabular}%
    }
  \label{tab:type}%
\end{table*}%

\textbf{\emph{Effectiveness on different assertion types.}} 
Table~\ref{tab:type} presents the performance of {\toolname} and baselines on different types of assertions.
We find that \toolname{} achieves optimal performance over all baselines on both datasets across all assertion types.
For the standard JUnit assertion types, \toolname{} achieves an accuracy of 58\%-73\% on \olddata{}, and 52\%-69\% on \newdata{}, outperforming \edit{} by 6.39\%-64.08\% and 11.05\%-33.84\%, respectively.
Particularly, for the \texttt{Equals} type, which is the most common standard type, \toolname{} generate 4993 and 6938 correct assertions on two datasets, 862 and 1599 more than \edit{}. 
Besides, for the non-standard assertion type (\ie \texttt{Other}), \toolname{} generates 1727 correct assertions with an accuracy of 58\% on \newdata{}, which are 251-717 (17.01\%-70.99\%) more than all baselines.
Overall, the results demonstrate \toolname{}'s generality to generate different types of assertions, including both standard and non-standard JUnit types.

\begin{figure}[t]
\centering
\includegraphics[width=0.9\columnwidth]{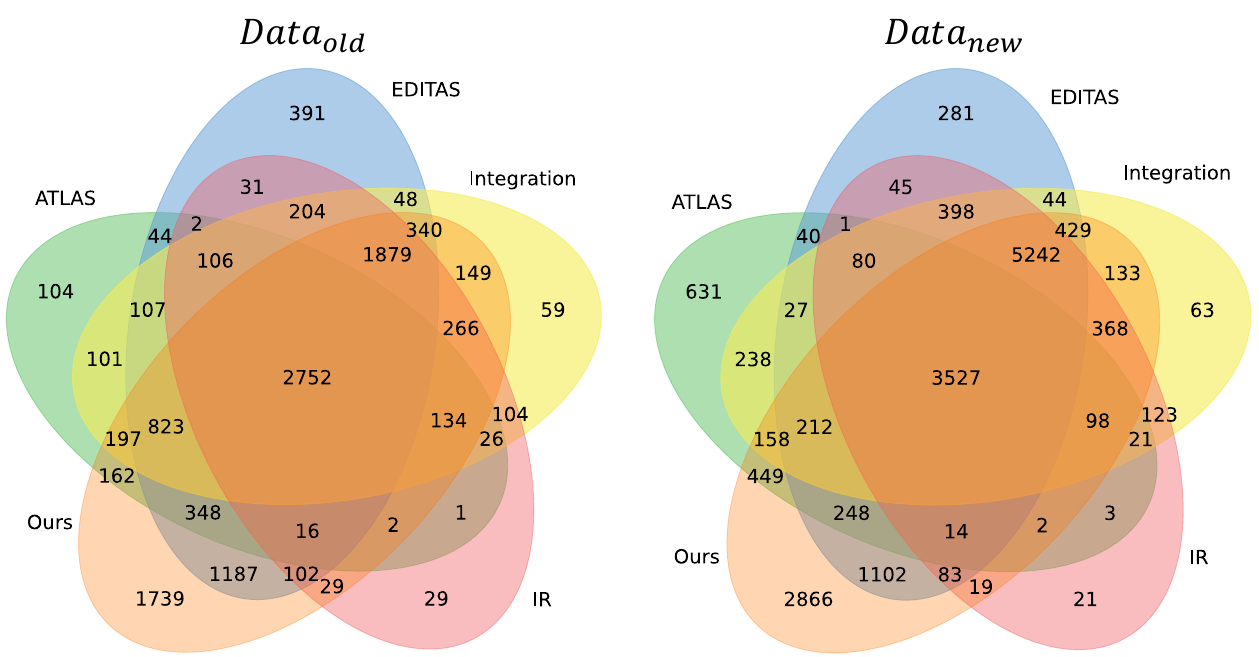}    
\caption{The overlaps of the unique generated assertions.}
\label{fig:veen}
\end{figure}

\textbf{\emph{Overlap Analysis.}}
~\figref{fig:veen} illustrates the number of unique assertions correctly predicted by \toolname{} and top-performing baselines in a Venn diagram format.
First, we find that \toolname{} successfully produces 1,739 accurate assertions on \olddata{} that all baselines fail to generate, which are 15.72X, 58.97X, 28.47X, and 3.45X more than \atla{}, \irar{}, \inte{}, and \edit{}, respectively.
Similarly, on \newdata{}, the improvement of \toolname{} is notably substantial with 2866 unique assertions, which are 3.54X, 135.48X, 44.49X, and 9.20X over the above four baselines.
Second, when particularly compared with the most recent baseline \edit{}, \toolname{} generates 2678 and 4093 unique assertions, outperforming \edit{} by 187.03\% and 346.83\% on both datasets, indicating the superior effectiveness of \toolname{} with the help of joint training.
Overall, our findings show \toolname{}'s ability to generate unique assertions, highlighting its potential to complement previous AG techniques.

\begin{figure}[t]
    \centering
    \includegraphics[width=0.9\linewidth]{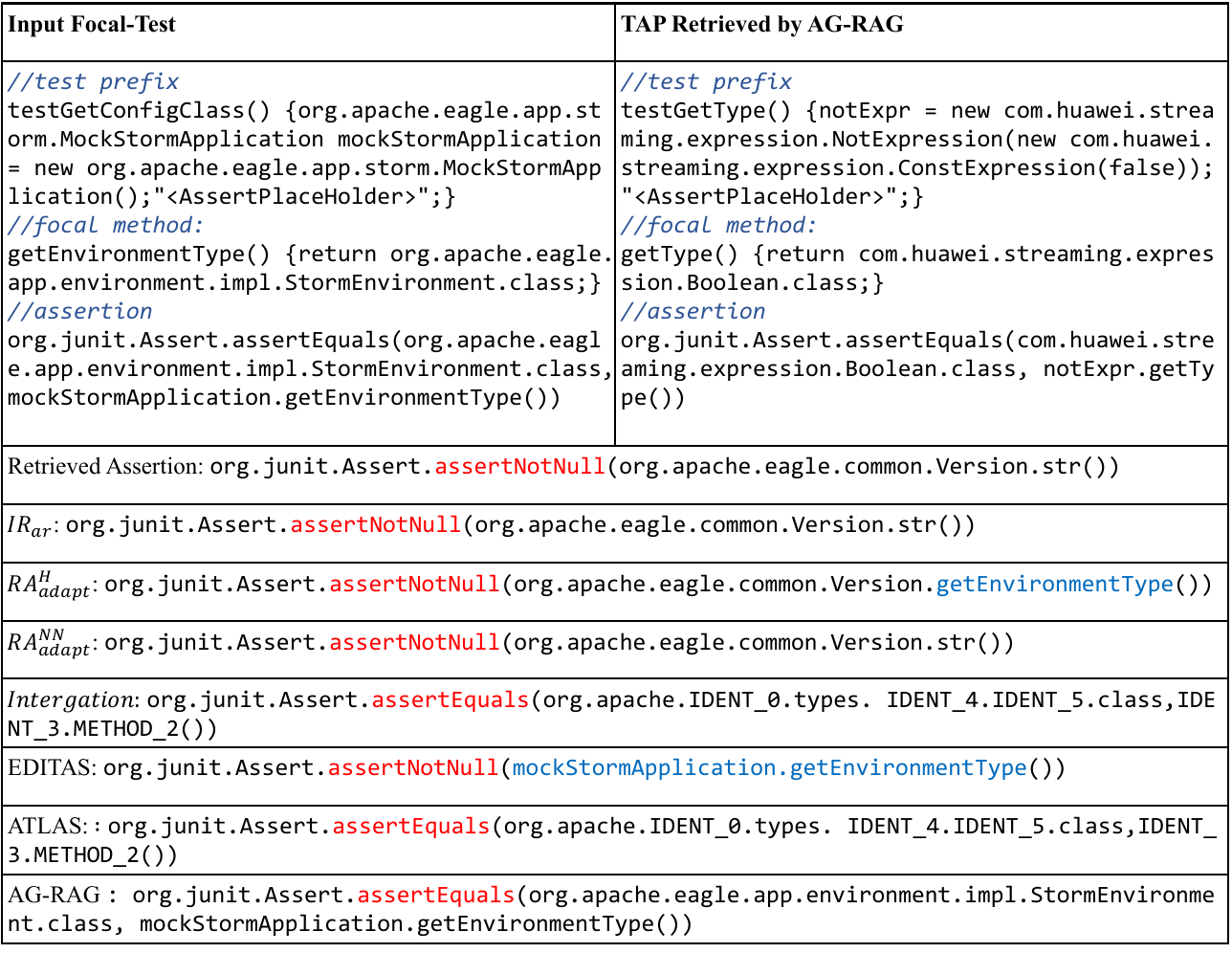}
    \caption{Example-1 of assertions generated by approaches}
    \label{fig:case}
\end{figure}

\begin{figure}[t]
    \centering
    \includegraphics[width=0.9\linewidth]{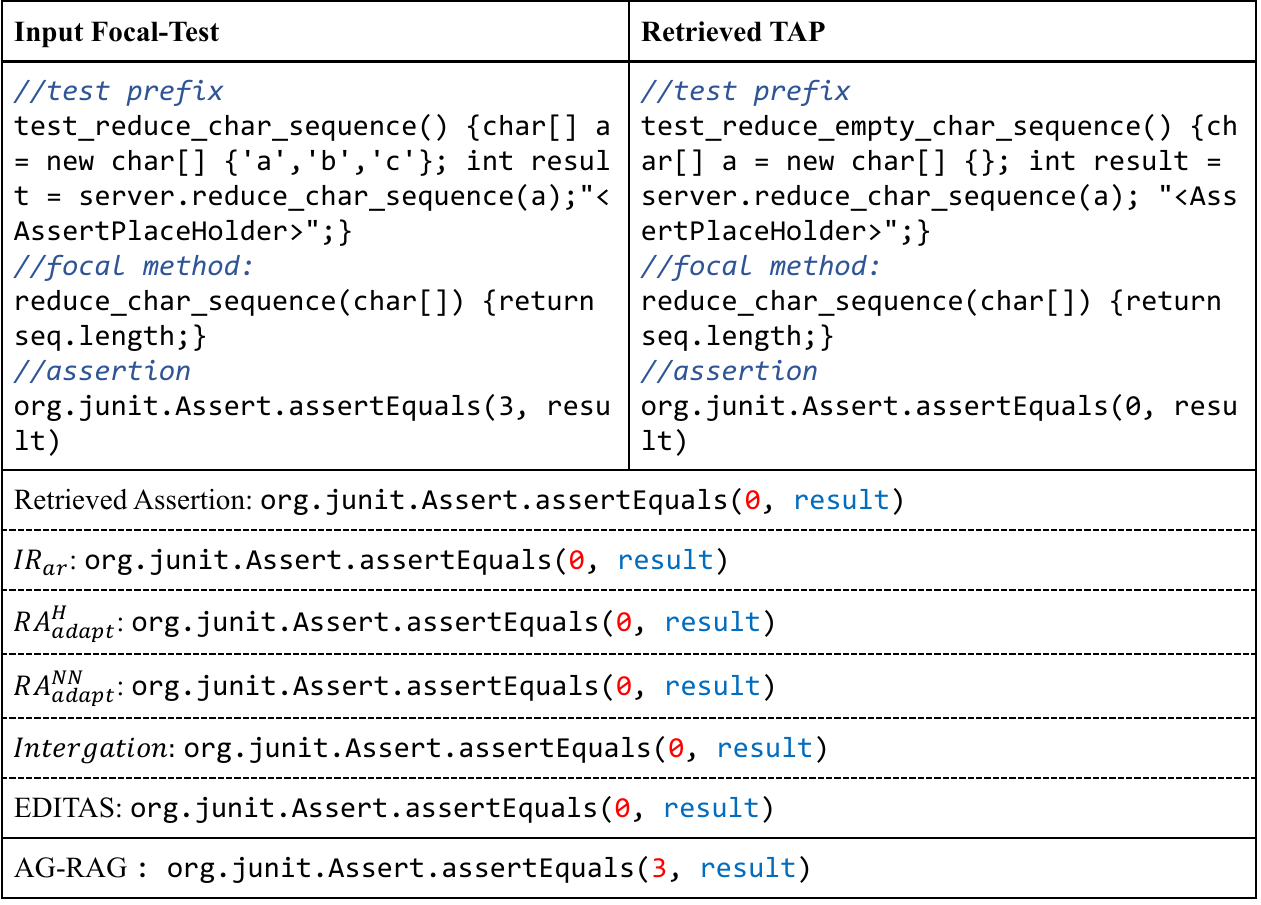}
    \caption{Example-2 of assertions generated by approaches}
    \label{fig:case2}
\end{figure}

\textbf{\emph{Case Study.}}
We present two examples to illustrate the retrieval and generation capabilities of \toolname{} respectively.
\figref{fig:case} illustrates an assertion example from the Apache Eagle project, which is only correctly generated by \toolname{}, but all baselines fail to.
In this example, \irar{} retrieves similar assertions based on lexical matching, and returns an assertion within the same project as the input focal-test.
Although the retrieved assertion has a high token similarity with that of the query focal-test (\eg both containing ``org.apache.eagle.common''), they are not responsible for testing similar functionalities.
Besides, \rah{}, \rann{} and \edit{} fail to produce correct assertions as all of them make modifications on the wrong assertion type, \ie \texttt{assertNotNull}.
For example, \rann{} attempts to replace the invoked function, and \edit{} chooses to replace the parameters within the assertion.
In contrast, {\toolname}, which relies on joint training, accurately retrieves a similar assertion from another project.
Despite significant differences in lexical matching, the two assertions share similar code semantics, \eg the same assertion type and parameter setting.
Thus, \toolname{} is able to capture the edit patterns between the two focal-tests, and performs the appropriate modifications on the retrieved assertion to generate the final correct assertion.
Similarly, another example can be seen in~\figref{fig:case2}, in which \toolname{} and all previous approaches retrieve the same assertion for the given focal-test.
The retrieved assertion is almost correct with only one parameter being refined ( ``0'' $\rightarrow$ ``3''), because it targets the same focal-method (``reduce\_char\_sequence'') as the ground truth.
However, existing baselines directly return this assertion, assuming it being correct.
\toolname{} successfully captures the semantic differences between the retrieved focal-test and the input focal-test (``char[] \{\}'' $\rightarrow$ ``char[] \{a, b, c\}''), and applies the corresponding edit operations to generate the correct assertion.

\begin{table*}[htbp]
  \centering
  \caption{Effects of the joint training in \toolname{}}
  \resizebox{0.85\linewidth}{!}{
  \begin{tabular}{c|ccc|ccc}
    \toprule
    \multirow{2}[4]{*}{\textbf{Appraoch}} & \multicolumn{3}{c|}{\olddata{}} & \multicolumn{3}{c}{\newdata{}} \\
\cmidrule{2-7}          & \textbf{Accuracy} & \textbf{CodeBLEU} & \textbf{BLEU} & \textbf{Accuracy} & \textbf{CodeBLEU} & \textbf{BLEU} \\
    \midrule
    No Retriever & 61.06\% ($\uparrow$5.78\%) & 77.36\% ($\uparrow$4.52\%) & 78.23\% ($\uparrow$8.39\%) & \multicolumn{1}{c|}{46.00\% ($\uparrow$22.45\%)} & 60.60\% ($\uparrow$12.39\%) & 62.02\% ($\uparrow$17.48\%) \\
    Random Retriever & 59.38\% ($\uparrow$8.77\%) & 77.22\% ($\uparrow$4.71\%) & 76.74\% ($\uparrow$10.49\%) & 44.12\% ($\uparrow$27.67\%) & 59.86\% ($\uparrow$13.78\%) & 58.86\% ($\uparrow$23.79\%) \\
    IR Retriever & 63.37\% ($\uparrow$1.93\%) & 78.54\% ($\uparrow$2.95\%) & 79.72\% ($\uparrow$6.36\%) & \multicolumn{1}{c|}{51.15\% ($\uparrow$10.13\%)} & 66.54\% ($\uparrow$2.36\%) & 67.41\% ($\uparrow$8.08\%) \\
    Pre-trained Retriever & 63.15\% ($\uparrow$2.27\%) & 78.99\% ($\uparrow$2.37\%) & 79.45\% ($\uparrow$6.72\%) & \multicolumn{1}{c|}{52.02\% ($\uparrow$8.29\%)} & 66.44\% ($\uparrow$2.51\%) & 67.44\% ($\uparrow$8.04\%) \\
    Fine-tuned Retriever & 64.19\% ($\uparrow$0.62\%) & 79.02\% ($\uparrow$2.33\%) & 80.19\% ($\uparrow$5.74\%) & \multicolumn{1}{c|}{53.40\% ($\uparrow$5.48\%)} & 67.96\% ($\uparrow$0.22\%) & 68.84\% ($\uparrow$5.84\%) \\
    \midrule
    Joint Retriever (\toolname{}) & \textbf{64.59\%} & \textbf{80.86\%} & \textbf{84.79\%} & \textbf{56.33\%} & \textbf{68.11\%} & \textbf{72.86\%} \\
    \bottomrule
    \end{tabular}%
    }
  \label{tab:rq2_retriever}%
\end{table*}%

\finding{1}{
{\toolname} significantly outperforms all prior AG approaches on three metrics, with a prediction accuracy of 56.33\%-64.59\% and 1739-2866 unique assertions on both datasets.
}

\subsection{RQ2: Analysis of Joint Training}
\label{sec:rq2}

\textbf{\emph{Experimental Design.}} 
In this section, we investigate how the proposed joint training strategy module affects performance in the retrieval-augmented generation setting.
We first consider a ``No retriever'' baseline that directly fine-tunes the generator with input focal-tests and their assertions in the training datasets.
We then compare our joint-training retriever in \toolname{} with different retrievers:
(1) a ``Random Retriever'' baseline that utilizes random sampling as the retriever;
(2) an ``IR Retriever'' baseline that utilizes IR as the retriever, following Yu~\etal~\cite{yu2022automated} and Sun~\etal~\cite{sun2023revisiting};
(3) a ``Pre-trained Retriever'' baseline that utilizes a pre-trained Codet5 without any fine-tuning as the retriever;
(4) a ``Fine-tuned Retriever'' baseline that first fine-tunes a pre-trained CodeT5 using the training set, and utilizes the trained CodeT5 as the retriever.

\textbf{\emph{Results.}}
Table~\ref{tab:rq2_retriever} presents the comparison results of our default retriever and baselines.
Overall, we find that the default joint retriever achieves the best performance on all metrics and datasets.
Particularly, IR Retriever achieves 57.26\% for accuracy, 72.54\% for CodeBLEU, and 73.57\% for BLEU on average, outperforming No Retriever by 7.48\%, 5.66\%, and 5.30\%, \revise{demonstrating the benefits of retrieving similar TAPs in guiding the assertion generation process and motivating \toolname{} to explore more powerful PLM-based retriever.}
Meanwhile, the downgraded performance of Random Retriever implies that randomly retrieved TAPs cannot provide helpful guiding signals \revise{due to the inherent noise in the randomly retrieved data, which lacks relevance}.
Besides, fine-tuning CodeT5 is able to retrieve more useful TAPs for the generator than the default CodeT5 model, with a prediction accuracy of 64.19\% and 53.40\% on both datasets.
\revise{The possible reason lies in that, compared with the default CodeT5, fine-tuning CodeT5 incorporates knowledge of assertion generation, which improves its ability to generate more effective embeddings for retrieval.}
Furthermore, the improvement of Joint Retriever against all baselines validates the effectiveness of our joint training module design, \revise{highlighting the substantial benefits of optimizing the retriever in conjunction with the generator, so as to retrieve useful TAPs.}

\finding{2}{
Our impact analysis demonstrates that our joint training strategy positively contributes to the performance of \toolname{} across three metrics, \eg improving No Retriever and Pre-trained Retriever by 22.45\% and 8.28\% accuracy on \olddata{}.
}

\subsection{RQ3: Generalizability of {\toolname}}
\label{sec:rq3}

\begin{table}[t]
  \centering
    \caption{Effectiveness of different PLMs in \toolname{}}
    \resizebox{0.9\linewidth}{!}{
    \begin{tabular}{c|cc|c}
    \toprule
    \textbf{PLMs} & \olddata{} & \newdata{} & Average \\
    \midrule
    GraphCodeBERT  & 54.58\% ($\uparrow$2.09\%) & 50.11\% ($\uparrow$12.96\%) & 52.34\% \\
    Unixcoder  & 59.54\% ($\uparrow$11.37\%) & 51.41\% ($\uparrow$15.89\%) & 55.47\% \\
    CodeBERT  & 55.22\% ($\uparrow$3.29\%) & 52.73\% ($\uparrow$18.86\%) & 53.97\% \\
    CodeT5 & \textbf{64.59\% ($\uparrow$20.82\%)} & \textbf{56.33\% ($\uparrow$26.97\%)} & \textbf{60.46\%} \\
    \midrule
    Average & 58.48\% ($\uparrow$9.39\%) & 52.64\% ($\uparrow$18.67\%) & 55.56\% \\
    \bottomrule
    \end{tabular}%
    }
  \label{tab:rq3_plms}
\end{table}%

\textbf{\emph{Experimental Design.}}
As mentioned in Section~\ref{sec_approach}, \toolname{} is a generic framework that can be easily integrated with different encoder-decoder Transformer PLMs.
To further investigate whether the performance of {\toolname} is affected by different PLMs, we consider three other advanced PLMs to replace CodeT5 in our framework: CodeBERT, GraphCodeBERT and UniXcoder.
All these PLMs are pre-trained with source code, publicly accessible, and medium-scale, thus suitable for fine-tuning in our work.

For encoder-only PLMs, like CodeBERT, we directly utilize them as the retrievers, and initialize a new decoder from scratch to construct an encoder-decoder architecture as the generators.
For encoder-decoder PLMs, like UniXcoder, similar to CodeT5 in Section~\ref{sec_approach}, we use their encoder part as the retrievers and the default encoder-decoder model as the generators.
It is noteworthy that we do not consider decoder-only PLMs, like CodeGPT~\cite{lu2021codexglue} and InCoder~\cite{fried2022incoder}, and even recent LLMs, such as Code Llama~\cite{roziere2023code} and StarCoder~\cite{li2023starcoder}, as they are built without an encoder, thus failing to provide meaningful code representations for assertion retrieving.

\textbf{\emph{Results.}}
Table~\ref{tab:rq3_plms} presents the comparison performance of \toolname{} with different PLMs as foundation models.
We only show the results of prediction accuracy due to page limit.
Overall, all PLMs consistently achieve impressive performance with an average accuracy of 58.48\% and 52.64\% on two datasets.
Particularly, we find all PLMs are able to achieve superior performance against previous AG approaches (shown in Table~\ref{tab:rq1}),
For example, five investigated PLMs achieve 54.58\%$\sim$64.59\% and 50.11\%$\sim$56.33\% prediction accuracy on \olddata{} and \newdata{}, improving the most competitive baseline \edit{} by 9.39\% and 18.67\% on average (highlighted as $\uparrow$ in each cell).
\revise{The substantial benefits demonstrate the generalizability of our framework, which can be integrated with other PLMs in a drop-in fashion.}
Besides, when comparing different PLMs, we find the default model of \toolname{} (\ie CodeT5) achieves better performance than the other three PLMs on two datasets.
For example, CodeT5 improves CodeBERT, GraphCodeBERT, and UniXcoder by 18.34\%, 8.49\%, and 16.97\% on the \olddata{} dataset.
\revise{Based on our analysis, we observe that the possible reason mainly lies in the model architecture.
CodeT5 is built on top of an encoder-decoder Transformer architecture, which is natural to support both the assertion retrieval (with an encoder) and generation tasks (with an encoder and decoder).
However, encoder-only models (\eg CodeBERT) require an additional decoder initialized from scratch to generate assertions, and have proven to be not suitable for code generation tasks~\cite{wang2021codet5}.
}

\finding{3}{
\toolname{} is general to different PLMs with an average accuracy of 58.48\% and 52.64\% on two datasets, and CodeT5 is remarkably effective in facilitating both assertion retrieval and generation with an accuracy of 64.59\% and 56.33\%.
}

\section{Discussion}

\begin{figure}[t]
	\graphicspath{{figs/}}
	\centering
	\includegraphics[width=0.98\linewidth]{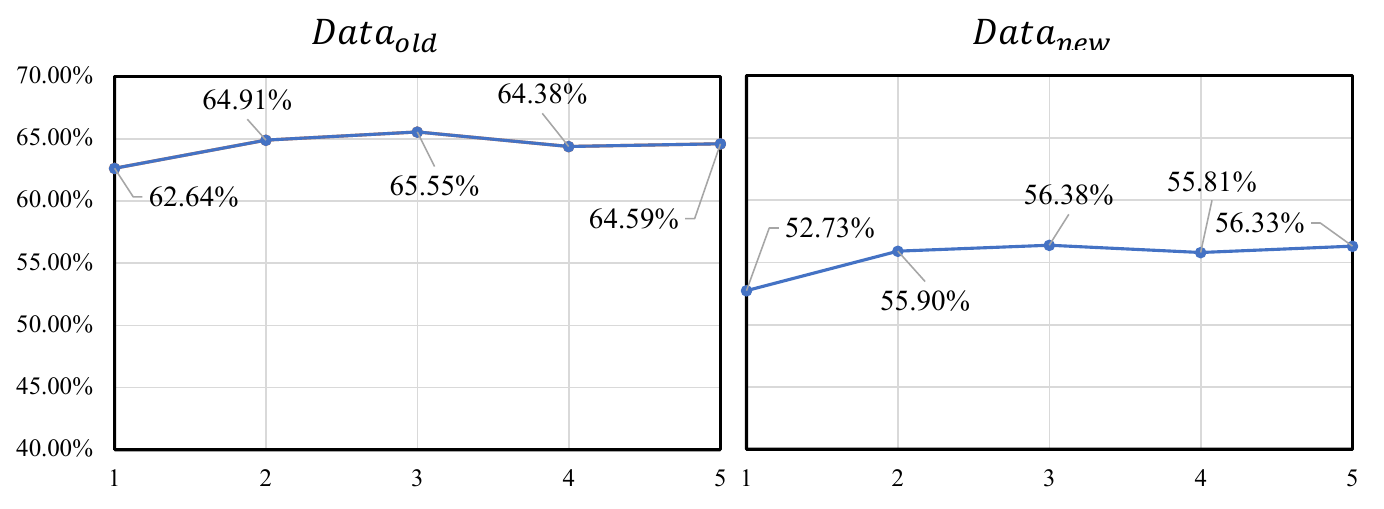}
	\caption{Effectiveness of the number of retrieved TAPs.}
	\label{fig:number_tap}
\end{figure}

\textbf{Analysis of the Number of Retrieved TAPs}.
As discussed in Section~\ref{sec:approach_joint}, we employ the top-$k$ retrieved TAPs with the highest probabilities to approximate the training objective of \toolname{}.
To explore how the number of retrieved TAPs affects the performance of \toolname{}, we set the minimal $k$ to one and the maximum $k$ to five due to resource limitations.
\figref{fig:number_tap} presents the prediction accuracy (Y-axis) of \toolname{} with different $k$ values (X-axis).
Overall, \toolname{} tends to achieve better performance with more retrieved TAPs.
Particularly, we find \toolname{} achieves the worst performance when $k$ is set to 1, with only 62.64\% and 52.73\% accuracy.
\revise{This is reasonable as, when $k=1$, only a single TAP is retrieved, with its retrieval probability consistently at 100\%, thus preventing the retriever from learning with the generator.}
We also find an accuracy increase of \toolname{} when $k$ is set from 1 to 3, \revise{suggesting that more retrieved TAPs provide more useful insights for generation, thereby facilitating more effective training.}
Furthermore, the performance of \toolname{} is similar when $k$ is between 4 and 5.
\revise{The possible reason is that more retrieved TAPs may lead to low-quality samples in the augmented inputs, thus affecting the training process.
Prior work~\cite{wang2023rap} has proven that irrelevant retrieved samples cannot provide useful guiding signals for the generator in program repair.
This motivates our choice of a pre-trained encoder as the retriever in Section~\ref{sec:approach_retrieval}, as it provides high-quality retrieved TAPs to guide the generation process.}

\textbf{Potential of Fault Detection Capabilities}.
As mentioned in Section~\ref{sec_metrics}, we utilize three static metrics to evaluate the performance of \toolname{} and baselines.
In this section, we attempt to explore the potential of generated assertions in uncovering real-world bugs.
Following prior work~\cite{dinella2022toga,liu2023towards,hossain2023neural}, we utilize EvoSuite~\cite{fraser2011evosuite} to generate test cases for Defects4J~\cite{just2014defects4j} that contains 835 bugs from 17 real-world Java projects.
We independently execute EvoSuite 10 times for each bug with different seeds to obtain the final test cases due to the randomized algorithms of EvoSuite.
We also exclude generated test cases involving exception behavior as our work focuses on the generation of assertions.
We then extract the test prefixes by removing assertion statements generated by EvoSuite, and leverage \toolname{} to predict a test assertion.
To identify the number of detected real-world bugs, we execute the complete test cases on both the buggy and the fixed program versions.
If a test case fails on the buggy version but passes
on the fixed one, it is considered capable of detecting the bug.
We compare \toolname{} against the best-performing baseline \edit{} due to the dynamic execution overhead and computational resources.
To ensure a fair comparison, we utilize \newdata{} as the training and retrieval corpus to implement both \toolname{} and \edit{}.
We find that \toolname{} and \edit{} detect 41 and 21 real-world bugs, of which 30 and 10 are undetectable by the other side of the hit.
\toolname{} outperforms \edit{} by 95.24\% and 100\% on the number of total and unique detected bugs.
This is the first attempt to evaluate fault detection capabilities of deep assertion approaches~\cite{sun2023revisiting,nashid2023retrieval,yu2022automated,watson2020learning}, and the results demonstrate the potential of \toolname{} in detecting bugs.
We will extend more experiments with real-world bugs in the future. 

\textbf{Potential of Large Language Models}.
Our work focuses on PLMs and selects CodeT5 to facilitate both assertion retrieval and generation tasks because CodeT5 is quite effective when fine-tuned to support code-related tasks.
We notice that recent LLMs have been released with powerful performance, such as CodeLlama~\cite{roziere2023code}.
Thus, we attempt to explore the preliminary potential of integrating LLMs with \toolname{}.
\revise{It is worth noting that black-box LLMs such as GPT-4o and GPT-3.5 are excluded because their models are not publicly released for fine-tuning.}
We select CodeLlama-7B as the foundation model to generate assertions and still leverage the CodeT5 Encoder to retrieve similar assertions, as CodeLlama is a decoder-only model without an encoder to generate representations.
Due to device limitations, we do not jointly train both the retriever and generator (\ie, retrieving assertions offline and only training the generator) and return one assertion for training.
We load CodeLlama in 8-bit quantization statistics~\cite{dettmers2022gptint} without full precision due to limited GPU memory and utilize parameter-efficient LowRank Adaption (LoRA)~\cite{hu2022lora} to train a small adapter (4M parameters) instead of the whole CodeLlama-7B with a reduction of 1600x.
To the best of our knowledge, this is the first attempt to fine-tune CodeLlama-7B in assertion generation research, representing the largest LLM in the community so far.
We find that CodeLlama achieves comparable performance against \toolname{} with an accuracy of 64.86\% on \newdata{}.
It is worth noting that the performance is valuable as we do not perform joint training, retrieve only one assertion, load CodeLlama without full precision, and train CodeLlama without full-parameter fine-tuning.
Despite PLMs being the focus of our work, the promising results motivate us to conduct more comprehensive experiments with newly released LLMs in the future.

\begin{table}[htbp]
  \centering
  \caption{\revise{Time Overhead of Joint Training in \toolname{}}}
    \begin{tabular}{c|cc|cc}
    \toprule
    \multirow{2}[4]{*}{\revise{\textbf{Epoch}}} & \multicolumn{2}{c|}{\revise{\textbf{CodeT5}}} & \multicolumn{2}{c}{\revise{\textbf{\toolname{}}}} \\
\cmidrule{2-5}          & \revise{\textbf{Time}} & \revise{\textbf{Accuracy}} & \revise{\textbf{Time}} & \revise{\textbf{Accuracy}} \\
    \midrule
    \revise{1} & \revise{1.84 h} & \revise{51.07\%} & \revise{2.9 h} & \revise{51.08\%} \\
    \revise{2} & \revise{3.69 h} & \revise{55.79\%} & \revise{5.82 h} & \revise{56.15\%} \\
    \revise{3} & \revise{5.55 h} & \revise{58.10\%} & \revise{8.74 h} & \revise{60.54\%} \\
    \revise{4} & \revise{7.41 h} & \revise{59.80\%} & \revise{11.66 h} & \revise{62.57\%} \\
    \revise{5} & \revise{9.27 h} & \revise{60.53\%} & \revise{14.57 h} & \revise{63.53\%} \\
    \revise{6} & \revise{11.12 h} & \revise{60.95\%} & \revise{17.49 h} & \revise{64.26\%} \\
    \revise{7} & \revise{12.99 h} & \revise{61.34\%} & \revise{20.41 h} & \revise{64.35\%} \\
    \revise{8} & \revise{14.84 h} & \revise{62.08\%} & \revise{23.34 h} & \revise{64.52\%} \\
    \revise{9} & \revise{16.71 h} & \revise{62.38\%} & \revise{26.28 h} & \revise{64.71\%} \\
    \revise{10} & \revise{18.59 h} & \revise{62.64\%} & \revise{29.22 h} & \revise{64.91\%} \\
    \bottomrule
    \end{tabular}%
  \label{tab:overhrad}%
\end{table}%

\revise{
\textbf{Efficiency of Joint Training}.
Following prior work~\cite{sun2023revisiting,yu2022automated}, we have demonstrated that \toolname{} achieves optimal performance across multiple evaluation metrics in Section~\ref{sec_results}.
In this section, we attempt to analyze the time overhead of joint training in \toolname{}.
To the best of our knowledge, this is the first attempt to explore the efficiency of deep assertion generation approaches in the community.
Table~\ref{tab:overhrad} presents the training time per epoch for AG-RAG (our joint training approach) and CodeT5 (\ie the baseline that direct fine-tunes CodeT5-based generator without joint training), along with the corresponding accuracy on \olddata{}.
We find that joint training incurs a higher training time compared to direct fine-tuning per epoch, which is reasonable as it requires optimizing both the retriever and the generator simultaneously. 
However, we further observe that this computational cost translates into significant gains in effectiveness.
\newdelete{For example, at Epoch 2, AG-RAG takes 5.82 hours and achieves an accuracy of 58.15\%, whereas the baseline, although requiring only 3.69 hours, resulted in a mere 55.79\% accuracy.}
\newrevise{For example, at Epoch 2, AG-RAG achieves an accuracy of 56.15\% in 5.82 hours, while the baseline, though faster at 3.69 hours, attains only 55.79\% accuracy.}
More importantly, when considering total time constraints, joint training demonstrates superior efficiency by achieving faster performance improvements. 
For example, AG-RAG reaches an accuracy of 62.57\% in 11.66 hours (Epoch 4), whereas the baseline requires 18.59 hours (Epoch 10) to achieve a comparable accuracy of 62.64\%.
Overall, while joint training introduces additional time overhead per epoch, it achieves faster convergence and better overall performance within the same computational budget.
These results highlight the efficiency and effectiveness of \toolname{}'s joint training framework, making it a compelling choice for the deep assertion generation task.
}

\begin{table}[htbp]
  \centering
  \caption{\revise{Impact of Sequence Embedding}}
    \begin{tabular}{c|ccc}
    \toprule
    \revise{\textbf{Embedding} } & \revise{\textbf{Accuracy}} & \revise{\textbf{CodeBLEU}} & \revise{\textbf{BLEU}} \\
    \midrule
    \revise{CLS Pooling} & \revise{64.59\%} & \revise{80.86\%} & \revise{84.79\%} \\
    \revise{Mean Pooling} & \revise{62.45\%} & \revise{78.71\%} & \revise{83.22\%} \\
    \revise{Max Pooling} & \revise{62.75\%} & \revise{77.84\%} & \revise{83.05\%} \\
    \bottomrule
    \end{tabular}%
  \label{tab:embedding}%
\end{table}%

\revise{
\textbf{Impact of Sequence Representation}.
As mentioned in Section~\ref{sec_approach}, \toolname{} utilizes the hidden state associated with the [CLS]  token as the input embedding.
The strategy is the common practice for PLMs to encode code snippets and the default sequence representation method in CodeT5, which serves as the foundational model for \toolname{}. 
To fully leverage CodeT5's pre-trained knowledge and code understanding capabilities, we directly adopt the vector corresponding to [CLS] as the sequence representation.
In this section, we attempt to explore how the embedding strategies influence the performance of \toolname{}.
We conduct an extended experiment to compare [CLS] with alternative aggregation strategies such as mean pooling and max pooling. 
Table~\ref{tab:embedding} presents the results of different representation strategies on the \olddata{} benchmark.
We can find the [CLS] strategy achieves the best performance across all evaluation metrics.
This confirms the appropriateness of our choice to use [CLS] as the sequence representation in this work.
Thus, we believe the use of the [CLS] strategy, as a standard practice in the community, has minimal influence on the validity of our results.
In the future, we recommend researchers perform an extensive and systemic investigation into the impact of different embedding aggregation methods in the retrieval-augmented deep assertion generation studies.
}

\revise{
\textbf{Test Generation vs. Assertion Generation}.
In the unit testing community, deep assertion generation studies (including \toolname{} and baselines) are motivated by the limitations of traditional test generation tools in capturing the intended program behavior with meaningful assertions.
Recently, deep learning, particularly PLMs, has shown promising potential in generating the whole test case~\cite{zhang2024testbench,wang2024hits, gu2024testart,schafer2023empirical,yuan2023no}.
In this section, we conduct an extended experiment to compare \toolname{} with end-to-end test generation with PLMs.
We consider fine-tuning CodeT5 with \olddata{} to generate test cases as a baseline, where the input is the focal method and the output consists of the corresponding test prefix and test assertion.
The results demonstrate that CodeT5 achieves only 3.06\% accuracy, 46.81\% BLEU and 51.87\% CodeBLEU, significantly underperforming \toolname{}.
These results are reasonable, as end-to-end test case generation demands a stronger capability for long-text comprehension, and longer outputs (consisting of both test prefixes and test assertions) significantly expand the search space—an issue that PLMs have long struggled to overcome.
Based on the results presented in Section~\ref{sec_results} and this section, we can find that PLMs excel at generating shorter test assertions by understanding focal-test semantics (\eg an accuracy of 64.59\% for \toolname{} on \olddata{}), but struggle with generating longer test prefixes (\eg only an accuracy of 3.06\%) that requires an understanding of the interactions between various functions.
In contrast, traditional test generation tools are capable of producing test prefixes with higher coverage through heuristic search (\eg EvoSuite~\cite{fraser2011evosuite}), but face challenges in understanding program semantics to generate meaningful assertions~\cite{almasi2017industrial}. 
Therefore, the two research areas are orthogonal, and 
at this stage, it is feasible and promising to combine the strengths of advanced PLMs with traditional test generation approaches.
Besides, generated assertions can not only complement existing test generation tools but also serve as code completion support for developers.
}

\section{Threats to validity}
\label{sec:threats}
\textbf{Internal Threat.}
The main internal threats are the potential of data leakage in \toolname{} and the selection of baselines.
\toolname{} is implemented with CodeT5, the pre-training samples of which may overlap with the testing samples of our benchmarks.
To address the concern, we carefully inspect CodeT5's pre-training datasets (\eg CodeSearchNet), and find it has not been exposed to any test cases, including assertions. 
It is worth noting that the data leakage concern motivates our choice of open-source PLMs, instead of more powerful black-box LLMs.
Thus, we confidently ensure that the pre-training data does not contain any overlap with the evaluation datasets in our experiments.
The second threat comes from the selection of baselines.
We exclude Mastropaolo~\etal~\cite{ mastropaolo2021studying, mastropaolo2022using} due to different research objects.
They pre-train a T5 model from scratch, while our work is built on top of off-the-shelf PLMs. 
Thus, in the future, similar to CodeT5, the work of Mastropaolo~\etal~\cite{ mastropaolo2021studying, mastropaolo2022using} can be utilized as the foundation model of \toolname{}.
We exclude CEDAR~\cite{nashid2023retrieval} due to the data leakage and reproducibility issues of its black-box LLM Codex, as mentioned in Section~\ref{sec:baselines}.
However, considering that \edit{} in RQ1 is the most recent AG technique and we include some PLMs as baselines (such as UniXcoder) in RQ2 and RQ3, the improvement of \toolname{} is enough to demonstrate the promising future in generating assertions by jointly fine-tuning the PLM-based generator and retriever.

\textbf{External Threats.}
The main external threat to validity lies in the utilized Java datasets.
The performance of \toolname{} may not extended to other programming languages.
However, \toolname{} is fully language-agnostic without considering any code-specific features, and can be applied to other languages directly.
Besides, Java is the most targeted language in the unit testing field due to the JUnit framework.
Furthermore, the two large-scale benchmarks are the most representative ones in deep assertion generation, and are adopted by all our baselines to yield reliable conclusions.
Thus, we believe that the impact of this threat is relatively minor to our conclusions.
In the future, we will explore the performance of \toolname{} on new benchmarks with more programming languages.

\textbf{Construct Validity.}
The main construct threat to validity comes from the evaluation metrics.
In our experiment, following all baselines~\cite{watson2020learning,yu2022automated,sun2023revisiting}, we evaluate the performance of \toolname{} with accuracy and BLEU due to the limitations of the utilized datasets.
We can not dynamically execute the whole program under test to determine whether generated assertions can detect real-world bugs, as the benchmark utilized only contains focal and test methods.
To address this threat, we additionally include CodeBLEU, a code-aware variant of BLEU that has not yet been adopted in prior AG work.
Besides, we introduce Defects4J to calculate the number of detected real-world bugs in a more realistic assessment setting.
In the future, we will evaluate the fault detection capabilities of generated assertions with more comprehensive benchmarks.

\section{Conclusion and Future Work}
\label{sec_conclusion}
In this paper, we propose a novel retrieval-augmented assertion generation approach, \toolname{}, by jointly training the retriever and generator with the help of external codebases and pre-trained language models (PLMs).
Given an input focal-test, \toolname{} first builds a dense retriever to search for relevant test-assert pairs (TAPs) with semantic similarity.
{\toolname} then utilizes off-the-shelf PLMs as the assertion generator to predict assertions with the input focal-test and augmented retrieved TAPs.
Besides, {\toolname} leverages a joint training strategy to optimize both the retriever and the generator with the retrieval probabilities and generation loss.
The experimental results on two widely-adopted datasets show the superior performance of \toolname{} against all six baselines on two metrics, \eg achieving 64.59\% and 56.33\% in terms of accuracy, outperforming all state-of-the-art AG techniques by 60.06\% and 56.72\% on average.
We also demonstrate that {\toolname} is able to generate a large number of unique assertions that all baselines fail to generate, \eg 1739 and 2866 on two datasets, 3.45X and 9.20X more than the most recent \edit{}.
We further demonstrate that {\toolname} is generalizable to different PLMs, consistently gaining superior performance against baselines.
In the future, we will further explore the applicability of \toolname{} with more powerful PLMs, benchmarks, programming languages, and dynamic metrics.

\section*{Acknowledgments}
We would like to thank the editors and anonymous reviewers for their time and comments.
This work is supported partially by the National Natural Science Foundation of China (61932012, U24A20337, 62372228), ShenzhenHong Kong-Macau Technology Research Programme (Type C) (Grant No. SGDX20230821091559018), the Fundamental Research Funds for the Central Universities (14380029), and the Open Project of State Key Laboratory for Novel Software Technology at Nanjing University (Grant No. KFKT2024B21).

\bibliographystyle{IEEEtran}
\bibliography{reference}

\end{document}